\documentclass[preprint,12pt,a4paper,nonatbib]{elsarticle}
\usepackage{amsfonts,mathrsfs, amssymb, amsthm}
\usepackage{amsmath}
\usepackage[ngerman,english]{babel} 
\usepackage{bm}
\usepackage{breakurl}
\usepackage{caption}
\usepackage{crimson}
\usepackage[utf8]{inputenc}
\usepackage{color,soul}
\usepackage{csquotes}
\usepackage{comment}
\usepackage{enumitem}
\usepackage{float}
\usepackage[T1]{fontenc}
\usepackage{graphicx}
\usepackage{hyperref}
\usepackage{MnSymbol}
\usepackage{mwe}
\usepackage{multirow}
\usepackage{parskip}
\usepackage{setspace}
\usepackage{subcaption}
\usepackage[super]{nth}
\usepackage{tabularx}
\usepackage{threeparttable}
\usepackage{url}

\makeatletter
\let\c@author\relax
\makeatother

\usepackage[
natbib      = true,
backend     = biber,
citestyle   = numeric,
bibstyle    = apa,
sorting     = none,		
url         = true,
isbn        = false,
doi         = true,
all dates   = year,
giveninits  = true,
]{biblatex}

\DeclareNameAlias{author}{last-first}

\DeclareFieldFormat[article,misc,inbook,book]{citetitle}{#1}
\DeclareFieldFormat[article,misc,inbook,book]{title}{#1} 

\AtEveryBibitem{
    \clearfield{urlyear}
    \clearfield{urlmonth}
    \clearlist{language}
}

\renewbibmacro{in:}{}

\DeclareFieldFormat{url}{\url{#1}}

\DeclareFieldFormat{doi}{%
  \ifhyperref
    {\href{https://doi.org/#1}{\nolinkurl{#1}}}
    {\nolinkurl{#1}}}

\DeclareFieldFormat{journaltitle}{#1\isdot}
\DeclareFieldFormat[article]{volume}{\apanum{#1}}

\makeatletter
\input{numeric.bbx}
\makeatother

\begin{document}

% ------------------------------------------------------------------
% ----- TITLE PAGE -------------------------------------------------
% ------------------------------------------------------------------

\begin{frontmatter}

% Title

\title{Geographical balancing of wind power decreases storage needs in a 100\% renewable European power sector}

% Wind power decreases the need for storage in an interconnected 100\% renewable European power sector
% Interconnection decreases storage needs in a 100\% renewable European power sector by balancing wind power variability
% Balancing the wind: why interconnection decreases storage needs in a 100\% renewable European power sector

% Authors
\author[1]{Alexander Roth \corref{cor}}
\ead{aroth@diw.de}
\cortext[cor]{Lead contact and corresponding author}

\author[1]{Wolf-Peter Schill}

\affiliation[1]{organization={German Institute for Economic Research (DIW Berlin)},
addressline={Mohrenstraße 58},
%postcode={},
city={10117 Berlin},
country={Germany}}

\begin{abstract}

To reduce greenhouse gas emissions, many countries plan to massively expand wind power and solar photovoltaic capacities. These variable renewable energy sources require additional flexibility in the power sector. Both geographical balancing enabled by interconnection and electricity storage can provide such flexibility. In a 100\% renewable energy scenario of twelve central European countries, we investigate how geographical balancing between countries reduces the need for electricity storage. Our principal contribution is to separate and quantify the different factors at play. Applying a capacity expansion model and a factorization method, we disentangle the effect of interconnection on optimal storage capacities through distinct factors: differences in countries' solar PV and wind power availability patterns, load profiles, as well as hydropower and bioenergy capacity portfolios. Results show that interconnection reduces storage needs by around 30\% in contrast to a scenario without interconnection. Differences in wind power profiles between countries explain around 80\% of that effect.

\end{abstract}

\begin{keyword}
variable renewable energy sources \sep electricity storage \sep interconnection \sep numerical optimization \sep 100\% renewable energy

\vspace{.2cm}

\JEL C61 \sep Q42 \sep Q49

\end{keyword}

\end{frontmatter}

% ----- New bib ---------------------------------------------------

%\begin{bibunit}
\begin{refsection}

% ----- Start doc --------------------------------------------------

\onehalfspacing

% --------------------
\section{Introduction}
% --------------------

The massive expansion of renewable energy sources is a major strategy to mitigate greenhouse gas emissions.\supercite{ipcc.2022} Thus, many countries have ambitious targets for increasing renewable shares in their power sectors.\supercite{ren21.2022} For example, the G7 countries aim for ``achieving a fully or predominantly decarbonized power sector by 2035''.\supercite{g7.2022} As the potentials for firm renewable generation technologies such as geothermal and bioenergy are limited in most countries, much of the projected growth needs to come from variable renewable energy sources, e.g.,~wind power and solar photovoltaics (PV).\supercite{child.2019} As these depend on weather conditions and daily and seasonal cycles, their electricity generation potential is variable.\supercite{lopezprol.2021} Increasing their share in the electricity supply thus requires additional flexibility of the power system to deal with their variability.\supercite{kondziella.2016} Geographical balancing, i.e.~transmission of electricity between different regions and countries, is a particularly relevant flexibility option.\supercite{schlachtberger.2017} This allows for balancing renewable variability over larger areas, using differences in load and generation patterns. Aside from such spatial flexibility, various temporal flexibility options can be used to manage the variability of wind and solar power, particularly different types of electricity storage.\supercite{schill.2020} Both geographical and temporal balancing can help to integrate surplus renewable generation and to meet residual load that could not be supplied by variable renewable sources at a particular location. 

From a techno-economic perspective, geographical balancing, using the electricity grid, and temporal balancing, using electricity storage, are substitutes for one another to a certain degree. Therefore, the need for storage capacities in a specific region decreases if electricity can be exchanged with neighboring areas that have partly uncorrelated weather and demand patterns. In an application to twelve central European countries, we investigate the interactions between geographical and temporal balancing, enabled by electricity storage, in a future 100\% renewable energy scenario. We do not aim to estimate the optimal amount of interconnection to be built; instead, we are interested in identifying and quantifying the drivers of why interconnection with neighboring countries mitigates electricity storage requirements. In terms of storage, we differentiate between ``short-duration'' storage, parameterized as lithium-ion batteries, and ``long-duration'' storage, parameterized as power-to-gas-to-power storage. We analyze the effects on both storage types separately. First, we measure the substitution effect between interconnection and storage by comparing the optimal storage capacities of two stylized least-cost power sector scenarios: in one electricity interconnection between countries is allowed; in the other, it is not. Then, we define several factors that can explain the reduced need for storage capacities in an interconnected electricity sector compared to one without interconnection. Finally, we quantify the magnitude of the different factors. 

We focus on five different factors to explain the storage-reducing effect of geographical balancing: differences between countries in hourly capacity factors of (1) wind and (2) solar power, which are a function of spatially heterogeneous weather patterns and daily and seasonal cycles; (3) hourly time series of the electric load; and the availability of specific technologies such as (4) hydropower and (5) bioenergy that differ due to geographic or historical factors. A capacity factor determines how much electricity a power plant can produce in a given hour compared to its installed capacity. E.g., a capacity factor of 50\% in a given hour means that a wind power plant with a power rating of 10~MW produces 5~MWh in that hour.

To determine the importance of each factor for storage capacity, we employ a factor separation method\supercite{stein.1993,lunt.2021}, which attributes model outcomes to different model inputs. This can be achieved by systematically varying only specific model inputs and comparing the outcomes of selected model runs. At the core of the analysis lies a comparison between an interconnected central European energy system with interconnection capacities foreseen by regulators\supercite{entso-e.2018} and a counterfactual system without any interconnection. The difference in optimal storage deployed by the model can be explained with the factor separation method.

To generate these model outcomes, we use an open-source model of the European electricity system that minimizes total system costs given an hourly exogenous electricity demand in each county. The model determines endogenously optimal investment and hourly usage of different generation and storage technologies for each country to meet the energy demand as well as other policy-related constraints, such as minimum-renewable requirements. Thus, market clearing is achieved every hour. The solution of a cost-minimizing model represents a long-run equilibrium in which, under idealized assumptions, all generators and storage assets exactly cover their fixed and variable costs with their revenues. The model comprises twelve central European countries that are connected in a ``net transfer capacity model'' with fixed interconnection capacities. For increased robustness, our analysis considers ten weather years from a 30-years period.

Several studies have estimated electricity storage needs in Europe in scenarios with high shares of renewables. Literature reviews identify a positive, linear relationship between renewable electricity shares and optimal electricity storage deployment.\supercite{cebulla.2018,blanco.2018} Focusing on single countries, such as Germany, various analyses find that storage needs depend on the model scope, e.g., on the number of sector coupling technologies included and on how detailed these are modeled, as well as on the availability of other flexibility options.\supercite{weitemeyer.2015, babrowski.2016, scholz.2017, schill.2018a} Other studies investigate how much storage is needed in the wider European power sector. While results again depend on model and technology assumptions, studies covering several European countries imply relatively lower storage needs than analyses focusing on a single country only.\supercite{bussar.2014, despres.2017, child.2019, moser.2020} Other analyses investigate the need for electricity storage in the U.S.\supercite{safaei.2015,desisternes.2016,phadke.2020} For instance, long-duration storage requirements in Texas increase with growing penetration of variable renewable energy sources.\supercite{johnson.2021} Related studies derive similar findings and also conclude that interconnection decreases storage needs, focusing on other parts of the U.S.\supercite{ziegler.2019} or the whole of the United States \supercite{tong.2020,dowling.2020,brown.2021, bloom.2022}. Similarly, geographical balancing and electricity storage are identified as partial substitutes in a model analysis of the North-East Asian region.\supercite{bogdanov.2016} This substitution is considered to be particularly relevant for long-duration storage technologies.\supercite{jenkins.2021} Various papers have analyzed wind and/or solar power variability and its impacts on the future energy system, partly focusing on extreme energy drought events.\supercite{collins.2018,raynaud.2018,cannon.2015,ohlendorf.2020,weber.2019} Yet, none of these studies focus primarily on quantifying the effect of interconnection on storage needs or on systematically isolating individual drivers of this effect.

Hence, we contribute to the literature by illustrating how spatial flexibility influences the need for temporal flexibility in an application to twelve central European countries. Our principal contribution is to quantify how different factors contribute to the reduction in storage capacity caused by geographical balancing. To identify the importance of these different factors, we use an adapted ``factor separation'' method.\supercite{stein.1993, lunt.2021} As there is so far no established method to attribute outcomes of power market models to changing model inputs, we propose a modified procedure that builds on counterfactual scenarios and factor separation, which could also be used in other energy modeling applications. We are the first to employ factor separation in the context of energy modeling, using it to quantify the importance of which factors drive down storage needs in an interconnected central European energy system.

% ------------------------------------
\section{Results} \label{sec: results}
% ------------------------------------

Employing a factor separation approach in combination with a numerical energy sector model, we determine by how much interconnection between countries decreases the overall optimal storage energy and power capacity of the energy system (Section \ref{subsec: results - main}). Afterward, we attribute the change in storage capacity to different drivers (Section \ref{subsec: results - drivers}) and explain the key mechanisms (Section \ref{subsec: results - explanations}).

\subsection{Geographical balancing reduces optimal storage power and energy capacity} \label{subsec: results - main}

\begin{figure}[H]

\centering

%     \begin{subfigure}[c]{0.45\textwidth}
%         \centering
%         \subcaption{Short-duration storage (energy)}
%         \includegraphics[width=\textwidth]{figures/1_agg_sto_energy_short.pdf}
%     \end{subfigure}
%     \begin{subfigure}[c]{0.45\textwidth}
%         \centering
%         \subcaption{Long-duration storage (energy)}
%         \includegraphics[width=\textwidth]{figures/1_agg_sto_energy_long.pdf}
%    \end{subfigure}
%    \vspace{4mm}
%    \begin{subfigure}[c]{0.45\textwidth}
%         \centering
%         \subcaption{Short-duration storage (discharging power)}
%         \includegraphics[width=\textwidth]{figures/1_agg_sto_power_short.pdf}
%    \end{subfigure}
%    \begin{subfigure}[c]{0.45\textwidth}
%        \centering
%        \subcaption{Long-duration storage (discharging power)}
%        \includegraphics[width=\textwidth]{figures/1_agg_sto_power_long.pdf}
%    \end{subfigure}

    \includegraphics[width=0.9\textwidth]{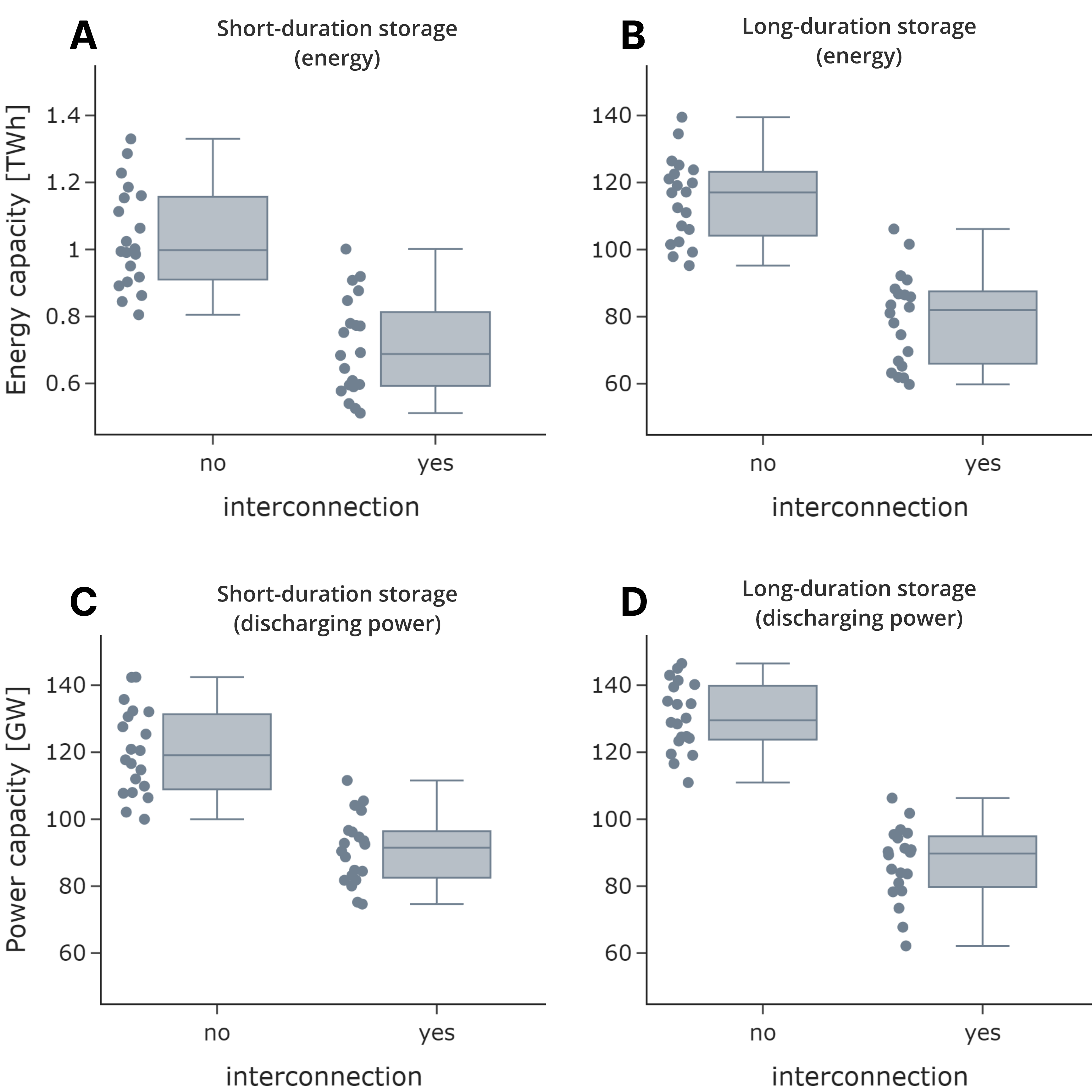}
    
    \vspace{1mm}
    \begin{minipage}[c]{0.90\textwidth}
    \medskip\footnotesize
        \emph{Notes:} The figure shows energy capacities (Panel A and B) and discharging power (Panel C and D) capacities of short- and long-duration storage aggregated over all countries. Every dot is the scenario result based on one weather year. The middle bar shows the median value. The box shows the interquartile range (IQR), which are all values between the 1st and 3rd quartile. The whiskers show the range of values beyond the IQR, with a maximum of 1,5 x IQR below the 1st quartile and above the 3rd quartile.
    \end{minipage}
     
    \caption{Aggregate installed storage energy and discharging power capacity}
    \label{fig: aggregate storage} 
\end{figure}

We find that aggregated optimal storage capacity is substantially lower in an interconnected system than in a system of isolated countries (Figure~\ref{fig: aggregate storage}). This applies to both short- and long-duration storage, as well as to storage discharging power and energy. Interconnection reduces optimal energy capacity need of short- and long-duration storage on average by 31\% over all years modeled. Discharging power, on average, decreases by 25\% for short-duration and by 33\% for long-duration storage. This translates to a reduction of 36~TWh in storage energy and 74~GW in storage discharging power (short- and long-duration storage combined) for the modeled interconnected central European power sector with 100\% renewable energy sources.

These results confirm previous findings in the literature that a system with interconnection requires less storage than a system without or put differently, that geographical balancing of variable renewable electricity generation across countries mitigates storage needs. We show that this also holds in a scenario with 100\% renewable energy. The variation of results between weather years is substantial, as optimal long-duration storage varies between 95 TWh and 140 TWh depending on th weather year. However,  our results indicate that the storage-reducing effect of interconnectinon is robust to different weather years.

\subsection{Wind power is the largest driver for mitigating storage needs}  \label{subsec: results - drivers}

\begin{figure}[H]

\centering

%\begin{subfigure}[c]{0.45\textwidth}
\centering
%\subcaption{Short-duration storage (energy)}
%\includegraphics[width=\textwidth]{figures/2_waterfall_energy_short.pdf}
%\label{subfig: waterfall a}
%\end{subfigure}
%\begin{subfigure}[c]{0.45\textwidth}
%   \centering
%   \subcaption{Long-duration storage (energy)}
%   \includegraphics[width=\textwidth]{figures/2_waterfall_energy_long.pdf}
%   \label{subfig: waterfall b}
%\end{subfigure}
%\vspace{1mm}
%\begin{subfigure}[c]{0.45\textwidth}
%     \centering
%     \subcaption{Short-duration storage (discharging power)}
%     \includegraphics[width=\textwidth]{figures/2_waterfall_power_short.pdf}
%     \label{subfig: waterfall c}
%\end{subfigure}
%\begin{subfigure}[c]{0.45\textwidth}
%    \centering
%    \subcaption{Long-duration storage (discharging power)}
%    \includegraphics[width=\textwidth]{figures/2_waterfall_power_long.pdf}
%    \label{subfig: waterfall d}
%\end{subfigure}
    \includegraphics[width=0.9\textwidth]{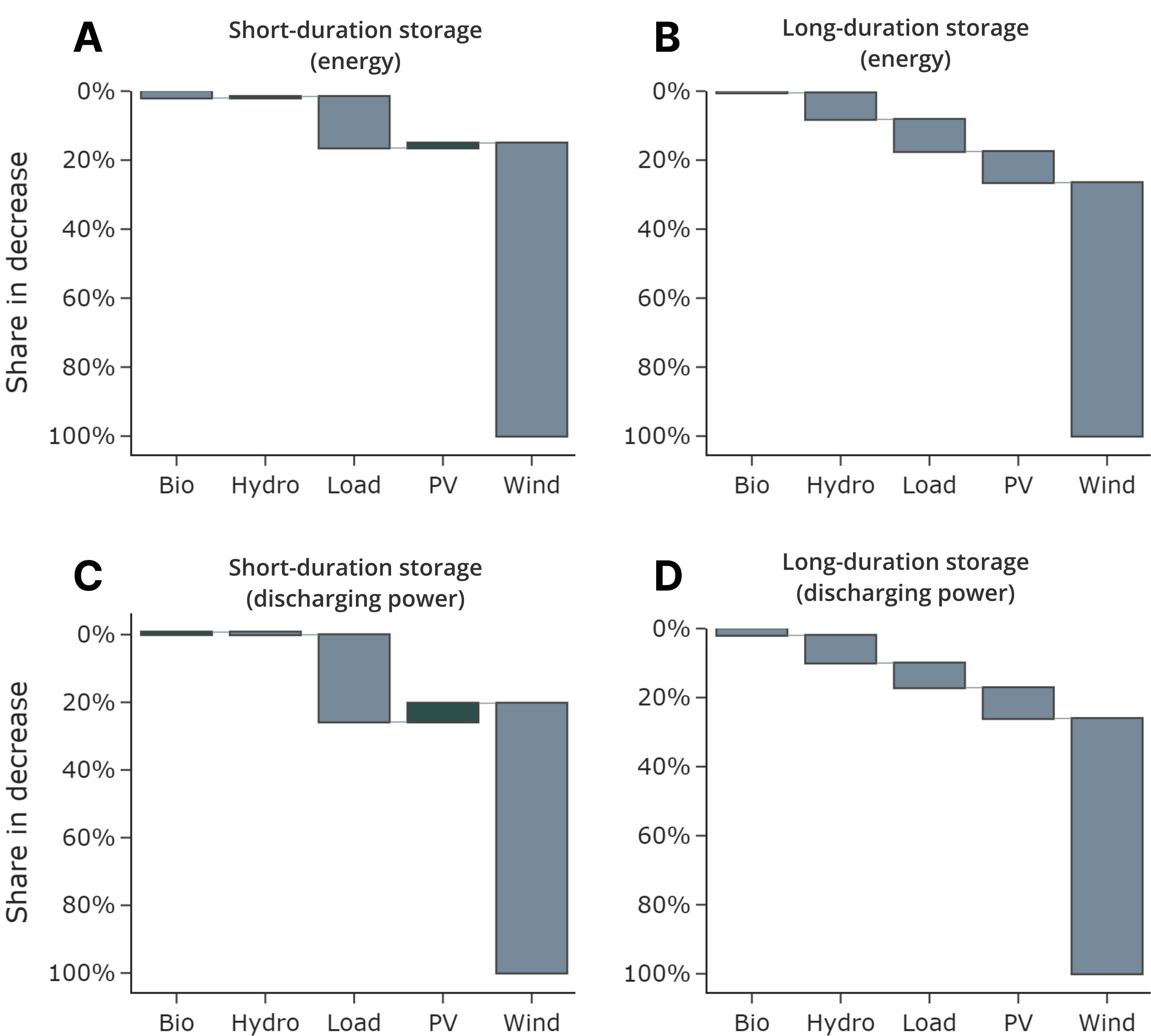}
    \vspace{1mm}
    \begin{minipage}[c]{0.90\textwidth}
    \medskip\footnotesize
        \emph{Notes:} The figure shows the average relative contributions of different factors to the reduction in storage energy (Panel A and B) and discharging power (Panel C and D) capacity due to interconnection. The average is taken over all ten weather years included in the analysis.
    \end{minipage}

\caption{Relative factor contribution to storage mitigation}
     
\label{fig: waterfall storage}
    
\end{figure}

Using counterfactual scenarios and a factorization method (more information in Section \ref{sec: factorization method}), we can attribute the decrease in optimal storage needs to individual factors. Wind power contributes by far the most, namely 80\%, to reducing storage discharging power and energy (Figure~\ref{fig: waterfall storage}). 

Especially for short-duration storage, differences in load profiles also contribute substantially to the storage-mitigating effect of interconnection. These account for 26\% of the decrease in short-duration storage discharging power (Figure~\ref{fig: waterfall storage}, Panel C). In contrast, differences in PV have, on average, a small increasing effect on short-duration storage energy and discharging power. However, this effect is strongly heterogeneous, depending on the weather year. For instance, solar PV can explain in some years up to 13\% of the drop in storage energy and 8\% of the drop in discharging capacity, yet in turn, has even a storage-increasing effect in other years (Figure \ref{fig: percentage change storage}). Allowing for transmission between countries may increase optimal overall PV investments, all other factors being constant and homogenized; this is because capacities grow in countries with higher PV full load hours, i.e.,~with lower PV costs. In turn, the need for short-duration storage then increases compared to a setting without transmission between countries because of higher diurnal fluctuations.

In the case of long-duration storage, all investigated factors contribute to the reduction of optimal storage investments enabled by interconnection. While wind power is again clearly dominating, differences in hydropower capacity, load curves, and PV time series almost equally contribute to reducing storage needs. 

While Figure~\ref{fig: waterfall storage} depicts average values, using ten weather years, results for individual years vary (see Figure~\ref{fig: percentage change storage} for more detail). Especially the contribution of wind power strongly differs between weather years. However, the relative contributions of the factors are qualitatively robust. In all analyzed weather years, we find that wind power is the dominating factor.

Figure~\ref{fig: waterfall storage} shows the already aggregated factors. In the supplemental information \ref{sec: si further results}, we provide further information on the magnitude of all factors from the factorization in all weather years (Figure~\ref{fig: all factors energy all years}) and in weather year 2016 (Figure~\ref{fig: all factors energy 2016}).

\subsection{An explanation of key mechanisms}  \label{subsec: results - explanations}

\begin{figure}[H]
    \centering
    %\begin{subfigure}[c]{0.3\textwidth}
    %     \centering
    %     \subcaption{Generation in the (combined) peak residual load hour}
    %     \includegraphics[width=\textwidth]{figures/power_peak_hour_generation.pdf}
    %     \label{subfig: generation peak hours}
    % \end{subfigure}
    % \hspace{2mm}
    % \begin{subfigure}[c]{0.6\textwidth}
    %     \centering
    %     \subcaption{Largest positive residual load events \\ \quad}
    %     \includegraphics[width=\textwidth]{figures/max_energy_events.pdf}
    %     \label{subfig: largest residual load events}
    %\end{subfigure}
    \includegraphics[width=0.9\textwidth]{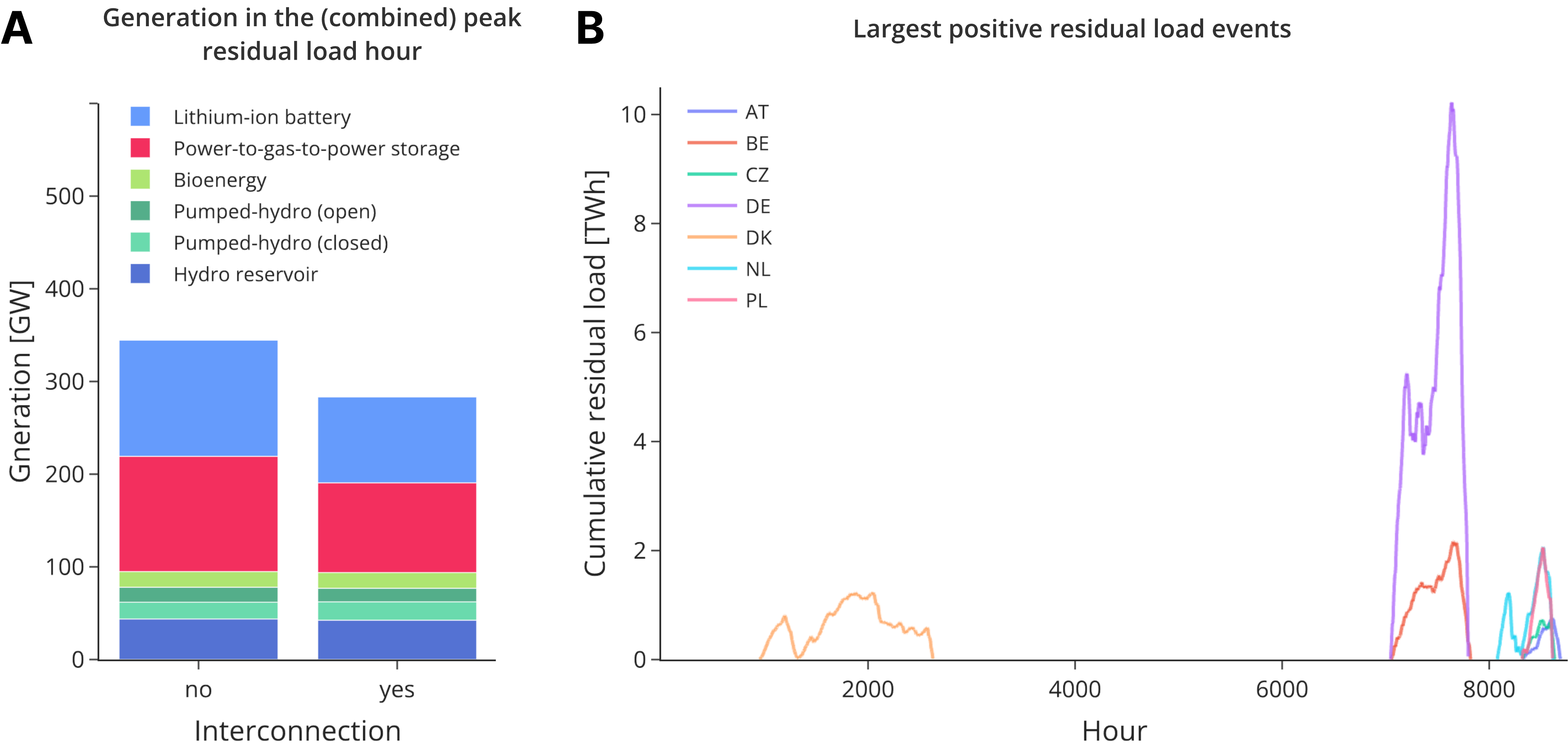}%
    \vspace{5mm}
    
    \begin{minipage}[c]{0.90\textwidth}
    \footnotesize
    \emph{Notes:} Panel A: The left bar shows the sum of electricity generation in the different countries' peak residual load hours, while the right bar shows the system-wide generation in the peak residual load hour of the interconnected system. Both bars depict the aggregate values of all countries. Panel B: Each country's largest positive residual load event is depicted. Countries with large hydro reservoirs are excluded as they have fundamentally different residual load events. Due to the existence of reservoirs, they accumulate large positive residual load events over the year. \\
    Both panels show data for the weather year 2016.
    \end{minipage}
    
    \caption{The drivers of reduced storage need: peak residual load hours and positive residual events}
    \label{fig: show drivers}
\end{figure}

To explain these results, we illustrate the key mechanisms using the weather year 2016. We turn to the peak residual load hour as a central driver to explain the drop in optimal storage discharging power capacity through interconnection. The peak residual load hour is defined as the hour in which residual load (i.e.,~load minus generation by variable renewable sources) is largest in a year. In an energy system based on 100\% renewables and high shares of wind and solar power, load in that critical hour has to be provided mainly by storage. Hence, the residual load peak hour determines the required storage discharging power capacity. 

When we compare an energy system without and with interconnection, the following thinking applies. In a system without interconnection, every country has to satisfy its own peak residual hourly load. Therefore, the overall (sum of all countries) storage discharging power needed in this system is simply the sum of all the countries' individual peak residual loads minus other existing generation options, such as bioenergy or hydro reservoirs. This simple addition is not true for an interconnected system if the countries' peak residual load hours do not coincide temporarily. Then, peak residual load hours in individual countries can potentially be compensated by geographical balancing, i.e., imports. Therefore,  the overall storage discharging power needed in an interconnected system is most likely smaller than the sum of the countries' peak residual loads.

The left bar of Figure \ref{fig: show drivers}, Panel A, shows the sum of electricity generation in the different countries' peak residual load hours, while the right bar shows the system-wide generation in the peak residual load hour of the interconnected system. The two differ because peak residual load hours do not align in the different countries. Implicitly, this reasoning assumes that there would be no limit on interconnection capacity between countries. In our case, net transfer capacities (NTC) are limited, so the residual peaks cannot be balanced out completely. Yet, even with limited interconnection, the non-aligned peak residual load hours of the different countries help to reduce residual storage discharging power needs.

To explain the reduced need for storage energy, a similar reasoning applies. The size of needed storage energy is correlated to the largest positive residual load event. We define a positive residual load event as a series of consecutive hours in which the cumulative residual load stays above zero. It may be interrupted by hours of negative residual load as long as the cumulative negative residual load does not outweigh the positive one. As soon as it does, the positive residual load event is terminated. These events typically occur when sunshine and wind are absent for long periods. 

An energy system with interconnection needs less storage energy if the countries' largest positive residual load events do not fully coincide. In this case, geographical balancing can help to flatten out these events. On the contrary, in a system without interconnection, all these events have to be covered in and by each country individually; hence the aggregate storage energy needs in a system without interconnection is the sum of every country's largest positive residual load event, and, therefore, higher than in a system with interconnection. Figure \ref{fig: show drivers}, Panel B, depicts the large positive residual load events for the year 2016 for different countries. Although some events overlap between the countries, many do not, and thus, interconnection helps reduce the need for storage energy capacity. 

\begin{figure}[!tb]
\centering
    %\begin{subfigure}[c]{0.45\textwidth}
    %    \centering
    %     \subcaption{Other countries' capacity factors \\ in peak residual load hours}
    %     \includegraphics[width=\textwidth]{figures/power_peak_hour_other_capa_factors.pdf}
    %     \label{subfig: capa factors}
    %\end{subfigure}
    %\begin{subfigure}[c]{0.45\textwidth}
    %     \centering
    %     \subcaption{Other countries' relative load \\ in peak residual load hours}
    %     \includegraphics[width=\textwidth]{figures/power_peak_hour_other_rel_load.pdf}
    %     \label{subfig: load}
    %\end{subfigure}
     \includegraphics[width=0.9\textwidth]{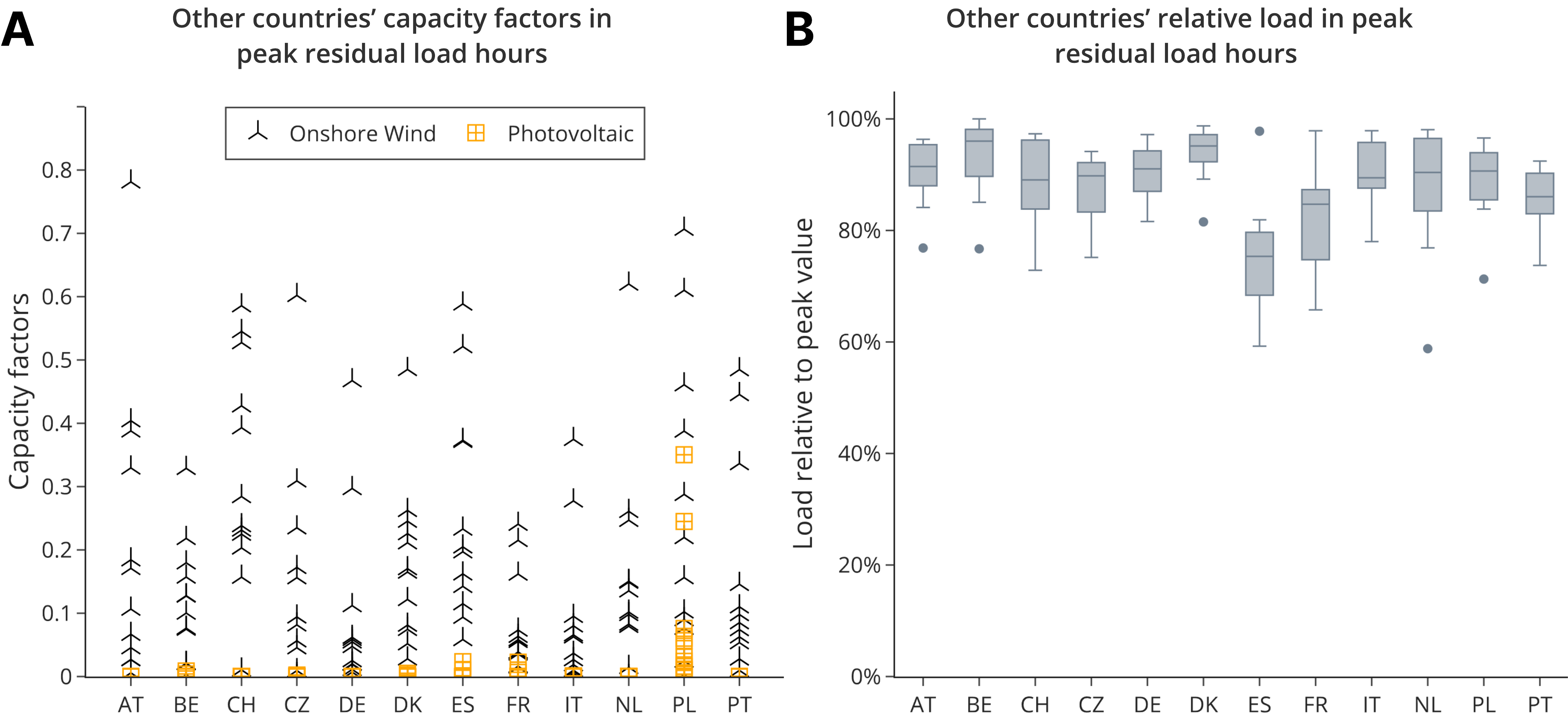}   
     \begin{minipage}[c]{0.90\textwidth}
     \medskip\footnotesize
        \emph{Notes:} Panel A shows hourly capacity factors of all other countries in the peak residual load hour of the country shown on the horizontal axis. Panel B shows the range of relative loads of all other countries in the peak residual load hour of the country shown on the horizontal axis. The middle bar shows the median value. The box shows the interquartile range (IQR), which are all values between the \nth{1} and \nth{3} quartile. The whiskers show the range of values beyond the IQR, with a maximum of 1,5 x IQR below the \nth{1} quartile and above the \nth{3} quartile. \\
        Both panels show data of the weather year 2016.
     \end{minipage}
     
    \caption{Illustration of main drivers: wind power, PV, and load}
    \label{fig: illustration of main drivers}
\end{figure}

As shown in the previous section, wind power is the principal factor that drives down storage needs when interconnection between countries is possible. Peak residual load and the largest positive residual load event largely determine storage needs. Therefore, the decrease in peak residual load and also in the largest residual load events are largely driven by the heterogeneity of wind power between countries. This can be confirmed in the data. In the hour of a country's highest residual load, onshore wind power capacity factors of most countries are still relatively high, so geographical balancing could help to make use of them (Figure~\ref{fig: illustration of main drivers}, Panel A). In contrast, this is hardly the case for PV capacity factors. The peak residual load hour of most European countries is likely to be in the winter when demand is high, but PV feed-in is low. Thus, wind power can contribute more to covering the peak residual hour than PV.

Load profiles also differ to some extent, such that relatively lower loads in other countries in combination with transmission can help to relieve the peak demand in a given country. During a peak residual load hour in a given country, we show the load (not residual load) relative to its maximal value in that year (Figure~\ref{fig: illustration of main drivers}, Panel B). Most countries have to cover their own load and have limited space to provide electricity for export. Most values range above 80\%. Therefore, differences in load profiles provide a positive but limited flexibility potential related to (peak) residual load balancing using interconnection.

Hydropower, a combined factor of hydro reservoirs, pumped-hydro, and run-of-river, has only a limited influence on storage reduction through interconnection. It could, in general, be an important provider of flexibility to the system. Yet, the reason for its limited importance is that installed hydro capacities are not big enough to substantially reduce the need for storage power and energy capacity (see Table~\ref{tab: assumption installed capacities}). This result may change under the assumption that the capacity of hydropower could be extended far beyond current levels. Then, the factor hydropower could play a bigger role in geographic balancing. This is also true for bioenergy, which we do not discuss here explicitly due to its minor effect.

% -------------------------------------------
\section{Discussion} \label{sec: conclusion}
% -------------------------------------------

\subsection{Interconnection decreases storage needs}

Identifying future electricity storage needs is highly relevant for planning deeply decarbonized, 100\% renewable power systems. \supercite{brown.2018} Using an open-source numerical model, our results show that optimal electricity storage capacity in an application to twelve central European countries substantially decreases when interconnection between countries is allowed. Compared to a setting without interconnection, short- and long-duration storage energy capacity decreases by 31\%; storage discharging power, on average, declines by 25\% and 33\%, respectively. These values hold for an average of ten weather years, covering three decades of historical data. Our outcomes corroborate and extend findings in the previous literature and show that the storage-mitigating effect of geographical balancing also holds in a with 100\% renewable energy. Yet, we go a step further by also disentangling and quantifying how the mitigation of storage needs is driven by different factors. To do so, we employ a factorization approach used, for instance, in climate modeling. \supercite{lunt.2021} To the best of our knowledge, this is the first time such an approach is adapted to a quantitative power sector model analysis. 

\subsection{Wind power is the most important factor}

We find that wind power is by far the most important factor in reducing optimal storage needs through geographical balancing. Its heterogeneity between countries accounts, on average, for around 80\% of reductions in storage energy and discharging power capacity needs. The main reason is that during peak residual load hours of a given country, which largely determine electricity storage needs, wind power availability in neighboring countries is still relatively high. Accordingly, geographical balancing helps to make better use of unevenly distributed wind generation potentials in an interconnected system during such periods. Differences in the profiles of solar PV and load, as well as in power plant portfolios (hydropower and bioenergy), contribute to the mitigation of storage needs to a much smaller extent. Though our analysis focuses on central Europe, we expect that qualitatively similar findings could also be derived for other non-island countries in temperate climate zones where wind power plays an important role in the energy mix.

\subsection{Conclusions on geographical balancing and modeling}

Our analysis fosters the grasp of the benefits of geographical balancing and its drivers. The findings may also be useful for energy system planners and policymakers. We reiterate the benefits of the European interconnection and argue that strengthening it should stay an energy policy priority if a potential shortage of long-duration electricity storage is a concern. Then, policymakers and system planners may particularly focus on such interconnection projects that facilitate the integration of wind power.

Finally, some modeling-related conclusions can be drawn. Any model analysis where wind power plays a role should properly consider geographical balancing in case storage capacities are of interest. Our analysis also indicates the importance of using more than one weather year in energy modeling with high shares of variable renewables. Not least, we hope to inspire other researchers to use factorization methods in energy modeling applications more widely. 

% -------------------------------------------
\subsection{Limitations}
% -------------------------------------------

As with any numerical analysis, our investigation comes with some limitations. First, we may underestimate storage needs due to averaging over specific weather years. In real-world systems, planners may pick only scenarios with the highest storage need to derive robust storage capacity needs. Likewise, planners may also want to consider an extreme renewable energy drought for storage dimensioning, i.e.,~a period with low wind and solar availability. In case such a renewable energy drought similarly affects all countries of an interconnection, the storage-mitigating effects may decrease. Second, we exclude demand-side flexibility options. In particular, we do not consider future sector coupling technologies such as battery-electric vehicles or heat pumps, which may induce substantial additional electricity demand but possibly also new flexibility options. Temporally inflexible sector coupling options may substantially increase storage needs.\supercite{schill.2020} Thus, we might overestimate the role of interconnection in mitigating storage. The interaction of sector coupling with storage mitigation via geographical balancing appears to be a promising area for future research. Third, optimization model results depend on input parameter assumptions. In particular, we assume fixed interconnection capacities (Table~\ref{tab: assumption installed ntc capacities}) and do not aim to determine the optimal amount of interconnection capacity investments. For such an analysis, a more detailed network model that considers optimal power flows over individual lines should be used. Larger interconnection capacities than assumed here could increase the storage-mitigating effect of interconnection as additional flexibility from other countries would become available. We show average utilization rates of interconnections in Section~\ref{sec: si further results}. Moreover, our analysis does not differentiate between the ``level'' and ``pattern'' effects of wind and solar PV profiles. In our counterfactual scenarios, we implicitly change both the patterns and the levels of wind and solar PV availability. Further analysis could disentangle these two factors and quantify this relative importance to better understand what exactly drives storage mitigation through wind and solar PV.

% ------------------------------------------------------------------
\section{STAR $\star$ Methods} \label{sec: methods}
% ------------------------------------------------------------------

\subsection{Key resources table}

\begin{table}[H]
\begin{tabularx}{\textwidth}{llX}
\textbf{RESOURCE} & \textbf{SOURCE} & \textbf{IDENTIFIER}   \\ \hline
\multicolumn{3}{l}{\textbf{Deposited data}} \\ \hline
Model and data & Gitlab & \url{https://gitlab.com/diw-evu/projects/storage_interconnection} \\ \hline
\multicolumn{3}{l}{\textbf{Software and algorithms}} \\ \hline
Python 3.8.16 & Python Software Foundation  &  \url{https://www.python.org/downloads/} \\
GAMS 36.1.0   & GAMS Development Corp	    & \url{https://www.gams.com/}  \\
DIETERpy 1.6.1 & Gaete-Morales (2021)\supercite{gaete-morales.2021} &  \url{https://gitlab.com/diw-evu/dieter_public/dieterpy/-/tags/v1.6.1}
\end{tabularx}
\end{table}

\subsection{Resource availability}

\subsubsection{Lead contact}

Further information and requests for resources should be directed to and will be fulfilled by the lead contact, Alexander Roth (\href{mailto:aroth@diw.de}{aroth@diw.de}).

\subsubsection{Materials availability}

Not applicable.

\subsubsection{Data and code availability}

\begin{itemize}
    \item Data: The data used in this paper can be accessed here: \url{https://gitlab.com/diw-evu/projects/storage_interconnection}.
    \item Code: The code used in this paper can be accessed here: \url{https://gitlab.com/diw-evu/projects/storage_interconnection}.
\end{itemize}

% ------------------------------------------------------------------
\subsection{Method details} \label{sec: experimental procedures}
% ------------------------------------------------------------------

\subsubsection{Factorisation method} \label{sec: factorization method}

Factorization (also known as ``factor separation'') is used to quantify the importance of different variables concerning their changes in a system. In complex systems, where more than one variable is altered simultaneously, it can be used to identify the importance of these variables for the changes in outcomes. Therefore, it can be used to analyze the results of numerical simulations.\supercite{lunt.2021}

There are several factorization methods, and our analysis builds on the factorization method by ``Stein and Alpert''\supercite{stein.1993} and its extension, the ``shared-interactions factorization''\supercite{lunt.2021}. The basic principle of factorization relies on comparing the results of various counterfactual scenarios to separate the influence of different factors on a specific outcome variable. For a broader introduction to factor separation, we refer to the Supplemental Information (\ref{sec: appendix factorization}) and to a recent paper\supercite{lunt.2021} providing an excellent introduction and overview.

To decompose the changes in storage needs, we define six factors that will impact the need for storage. Each factor can take two different states, which, to ease explanations, we call A and B. Table \ref{fig: factors and states} provides an overview of all factors and their possible states.

\begin{table}[H]
\centering
\begin{tabular}{l|ll}
Factor              & State A     & State B        \\ \hline
(1) Interconnection & not allowed & allowed        \\
(2) Wind            & harmonized  & not harmonized \\
(3) Solar PV        & harmonized  & not harmonized \\
(4) Load            & harmonized  & not harmonized \\
(5) Hydropower      & harmonized  & not harmonized \\
(6) Bioenergy       & harmonized  & not harmonized
\end{tabular}
\caption{Factors and states}
\label{fig: factors and states}
\end{table}

To determine the magnitude of the different factors, we compare model outcomes of different scenario runs. We compare a default ``real-world'' setting to a counterfactual setting. In the counterfactual setting, corresponding to state A, all factors are \textit{harmonized} which means that their respective cross-country variation is eliminated. In contrast, in the state B \textit{not harmonized}, all countries exhibit their own solar PV capacity factors. The same logic generally applies to the other factors as well. A more detailed definition and explanation of the factors is provided in the next section \ref{sec: factor definition}.

In contrast to other applications of factor separations, we are not interested in the \textit{entire} effect of each factor on storage needs. To identify which factors are most important in influencing storage needs through interconnection, we focus instead on the ``interaction terms'' between interconnection (1) and the other factors (2)-(6).

To identify the influence of the factors, we run several counterfactual scenarios. The notation to define the different factors is as follows. Whenever a factor is in state B, hence \textit{allowed} or \textit{not harmonized}, a subscript with the respective number is added. If the factor is in state A, no subscript 1-6 is added. The scenario in which all factors are in state A is called $f_0$, hence all factors are \textit{harmonized}, and no interconnection is allowed. In this scenario, all modeled countries are very similar, i.e., they have the same capacity factors, load patterns, and equal relative installed hydropower and bioenergy capacities. The scenario $f_1$ is nearly identical, with the expectation that interconnection is allowed as it is indicated by subscript 1, pointing to the factor interconnection. Following that logic, scenario $f_2$ resembles $f_0$, except that factor (2), i.e.,~wind, is not harmonized. Following that structure, we can define and name all relevant scenarios. For instance, $f_{12}$ denotes the scenario in which interconnection is allowed, and wind capacity factors are not harmonized, yet all the other factors are in their state A, hence \textit{harmonized}.

Of all possible scenarios, two are of special interest:

\begin{itemize}
    \item $f_{123456}$: This scenario can be regarded as our ``default'' scenario with no capacity factors or power plant portfolios being harmonized and interconnection between countries allowed.
    \item $f_{23456}$: This scenario equals the previous one, with the only difference that interconnection between countries is not allowed. Thus, all countries operate as electric islands.
\end{itemize}

We aim to explain the difference in optimal storage energy and power installed between these two scenarios $f_{123456}$ and $f_{23456}$, and to attribute the difference to the various factors (2)-(6). To quantify the importance of the different factors, we calculate the size of interaction factors between factor \textit{interconnection (1)} and the other factors (2)-(6).

The size of the individual factors can be defined as differences between scenario runs. These are denoted $\hat{f}_1$, $\hat{f}_2$, ..., $\hat{f}_{12}$, ..., etc. $\hat{f}_1$ is the sole effect of factor (1) by comparing the scenarios $f_0$ and $f_{1}$: 
\begin{equation}
    \hat{f}_1 = f_1 - f_0.  
\end{equation}
As described above, we rely on the interaction effects of factors for our attribution. The definition of interaction effects is more complicated and requires the results of several scenarios. For instance, the combined effect of the factors (1), (2), and (3), denoted $\hat{f}_{123}$, is defined as:
\begin{equation}
    \hat{f}_{123} = f_{123} - (f_{12} + f_{13} + f_{23}) + (f_1 + f_2 + f_3) - f_0
\end{equation}
Put in words, $\hat{f}_{123}$ measures \textit{only} the \textit{combined} influence of the factors interconnection (1),  wind (2), and  PV (3) on storage need, hence the interaction effect. The (direct) effects of the factors (such as $\hat{f}_{1}$) are not comprised.

To quantify the importance of different factors of interconnection on storage, we first define the ``difference of interest'' (INT), which we define as:
\begin{equation}
    INT = f_{123456} - f_{23456}.
\end{equation}
Then, we quantify which factors explain most of this difference. INT can be written as the sum of all interaction factors between the different factors (2)-(6) and the interconnection factor (1). Hence, every element of that sum has to comprise at least factor (1). It can be shown that the difference $INT$ is the sum of all the interaction factors where interconnection is involved, therefore

\begin{align}
    \text{INT} = & \hat{f}_{1} + \hat{f}_{12} + \hat{f}_{13} + \dots + \hat{f}_{16} + \hat{f}_{123} + \dots +  \hat{f}_{156} \nonumber \\ 
    & + \hat{f}_{1234} + \dots + \hat{f}_{1456} + \hat{f}_{12345} + \dots + \hat{f}_{13456} + \hat{f}_{123456}.
\end{align}

To calculate the contribution of one of the factors on the difference of interest, $INT$, we collect all interaction effects between the factor interconnection (1) and the respective other factor. For instance, to quantify the contribution of the factor  wind (2), we sum up all interaction effects that include the factors interconnection (1) and wind (2). The principal interaction effect $\hat{f}_{12}$ is part of it, but, e.g., also the interaction effects between interconnection, wind, and PV: $\hat{f}_{123}$. To avoid double-counting, we have to distribute these shared interaction effects between - in this case - the factors wind and PV. There are different ways to distribute these effects. We use the ``shared-interactions factorization''\supercite{lunt.2021} that distributes the interaction effects equally between the different factors. Hence, the total interaction effect between the factors interconnection and wind can be defined as follows:
\begin{equation}
    \hat{f}_{12}^{total} = \hat{f}_{12} + \frac{1}{2} \hat{f}_{123} + \frac{1}{2} \hat{f}_{124} + ... + \frac{1}{3} \hat{f}_{1234} + ... + \frac{1}{5} \hat{f}_{123456}
\end{equation}
Similarly, we define the interactions between interconnection and PV as $\hat{f}_{13}^{total}$, between interaction and load as $\hat{f}_{14}^{total}$, between interaction and hydropower as $\hat{f}_{15}^{total}$, and between interaction and bioenergy as $\hat{f}_{16}^{total}$. 

All these interaction terms $\hat{f}_{1i}^{total}$ add up to our difference of interest: 
\begin{equation}
    \text{INT} = \hat{f}_{12}^{total} + \hat{f}_{13}^{total} + \hat{f}_{14}^{total} + \hat{f}_{15}^{total} + \hat{f}_{16}^{total}.
\end{equation}
To determine the contribution of each factor (wind, PV, load, etc.) to the change in optimal storage capacities facilitated through interconnection, we calculate their share $s$. For instance, for the factor wind, this share reads as 
\begin{equation}
s_{wind} =\hat{f}_{12}^{total} / {INT}.
\end{equation}
As we have defined six factors, we need to run $2^6 = 64$ scenarios for a complete factorization of one weather year. As we perform our analysis for ten different weather years, we run 640 different scenarios (see Table~\ref{tab: appendix overview runs} for an illustrative overview).

\subsubsection{Definition of factors} \label{sec: factor definition}

The basic principle to quantify how different factors impact optimal storage through interconnection is the use of counterfactual scenarios, in which the state of these factors is varied. For all our factors, we define two states in which they can exist. For most of the factors, these states are \textit{not harmonized} and \textit{harmonized}, in which, in the latter, all countries are made equal to eliminate the variation between countries. By ``making equal'', we refer to a counterfactual scenario in which differences between countries, such as different renewable energy availability time series or hydropower availabilities, are eliminated.

We define five factors we consider to be most relevant. The two factors  ``wind'' and ``PV'', covering most of the energy supply, are associated with the variable capacity factors of these technologies. Another factor is ``load'' which covers energy demand. The two factors ``hydropower'' and ``bioenergy'' relate to different inherited power plant portfolios in different countries. Finally, the factor ``interconnection'' is defined only to make the analysis operational, not to explain reduced storage needs.

\paragraph{Wind} \label{sec: factor wind}

The factor that captures the impact of wind patterns is operationalized with the help of capacity factors and takes two different states: \textit{not harmonized} or \textit{harmonized}. In the state \textit{not harmonized}, every country has its own capacity factor time series, as provided by the database used \supercite{defelice.2020} (more information in Section \ref{sec: appendix assumptions and data}) given the specific weather year. On the contrary, in the state \textit{harmonized}, capacity factors are equal in all countries using the capacity factors of our reference country German. Hence, all variation between countries in wind power capacity factors is taken away.

On top, we also have to account for geographic differences in offshore wind power, which cannot be deployed in all countries because of differences in access to the sea. In contrast to onshore wind power and solar PV which could be, in principle, deployed everywhere, wind offshore, like hydropower, cannot. In the state \textit{harmonized}, not only do the capacity factors have to be the same across all countries, but also all countries have to operate ``as if they are the reference country'' (Germany in our case). Therefore, in the state \textit{harmonized}, all countries exhibit the same share of offshore wind power plants. That share is defined as installed capacity divided by the total yearly load. We use the total yearly load as the denominator as it is not related to the power plant fleet but is still country-specific. If we used the share of installed power plant capacity, the model would have the incentive to change the total power plant fleet, which we have to avoid. This share is determined based on a scenario run of our reference country Germany in isolation. 

Using this approach implies, given the share is larger than zero, that also countries without sea access, e.g., Austria or Switzerland, have offshore wind power plants in the state \textit{harmonized}. Although this is clearly not realistic, this harmonization step - including the application of the share - is necessary to take away all the cross-country variation of capacity factors, and also geographic differences such as access to the sea. In the state \textit{harmonized}, all countries act as if they were the reference country in isolation. 

\paragraph{Solar PV}

The factor \textit{solar PV}, like wind power, takes two states. The state \textit{not harmonized} corresponds to the default with solar PV capacity factors as provided by our data source. In the \textit{harmonized} case, PV capacity factors are equal in all countries using those of our reference country. Hence, all variation between countries in solar PV capacity factors is taken away.

\paragraph{Load} 

The definition of the factor ``load'' is similar to factors ``wind power'' and ``solar PV''. In the state \textit{harmonized}, all countries have the same load time series as our reference country, yet scaled to their original total yearly demand. Therefore, in the state \textit{harmonized}, all countries have the same load profile (same as the reference country Germany) but on country-specific levels.

\paragraph{Hydropower}

In addition to differences in wind, solar PV, and load patterns, we also aim to quantify how much of the storage capacity reduction can be attributed to specifics of the existing power plant portfolios because of legacy capacities and limited expansion potentials. Hydropower, comprising reservoirs, pumped-hydro, and run-of-river, can be considered to be exogenous. Some countries happen to have them while others do not. Also, their installed generation capacities are considered to be exogenous.

In the state \textit{harmonized}, all countries have the same share of installed power plant capacities of the respective technologies. We treat all countries as if they had a power plant portfolio like the reference country in isolation. In the case of hydropower, we also assume the German hydro times series for the other countries. These shares are determined based on a scenario run of our reference country Germany in isolation. We calculate the relative weight of the exogenous technologies as a share of installed capacity over the total yearly load. In the state \textit{harmonized}, this share is applied to all countries. For a detailed explanation regarding the shares, we refer to paragraph \textit{Wind} above.

\paragraph{Bioenergy}

The definition of the factor \textit{bioenergy} closely follows the one of hydropower described above. In the state \textit{harmonized}, all countries have the same share of installed bioenergy power plant capacities. We consider all countries as if they had a power plant portfolio like the reference country in isolation. 

\paragraph{Interconnection}

The factor interconnection is needed to make the factor separation operational. Like the other factors, it has only two states. In contrast to the other factors, they are called \textit{not allowed} and \textit{allowed} and determine whether electricity flows between countries is possible. In the state \textit{allowed}, interconnection is allowed and the interconnection capacities between countries are fixed, as given in Table \ref{tab: assumption installed ntc capacities}. If interconnection is \textit{not allowed}, electricity flows between countries are not possible.

\subsubsection{Model} \label{sec: model}

To obtain the model results needed for the factor separation, we use the open-source capacity expansion model DIETER\supercite{zerrahn.2017,gaete-morales.2021}, which has previously been used for detailed long-term electricity sector planning analyses \supercite{schill.2018a,say.2020,stockl.2021,gils.2022b,vanouwerkerk.2022} and for more stylized illustrations\supercite{schill.2020,kittel.2022,gils.2022a}. It minimizes total power sector costs for one year, considering all 8760 consecutive hours. DIETER focuses on the temporal flexibility of renewable integration. It assumes unconstrained electricity flows within countries. In this application, the model comprises 12 central European countries: Austria, Belgium, Czechia, Denmark, France, Germany, Italy, Netherlands, Poland, Portugal, Spain, and Switzerland  (Figure~\ref{fig: countries}). In scenarios in which electricity exchange between countries is allowed, countries are connected with a transport model based on Net Transfer Capacity (NTC). These are fixed according to an ENTSO-E scenario (Table~\ref{tab: assumption installed ntc capacities}); an expansion or reduction of these cross-border interconnection capacities is not possible. The model does not consider transmission or distribution bottlenecks within a country. \sloppy

Endogenous model variables of interest are the installed capacity of on- and offshore wind power and solar PV and the installed capacity of short- and long-duration storage, differentiated by storage energy, as well as charging and discharging power. Further model outputs are hourly patterns of electricity generation and curtailment (of renewables), the charging and discharging patterns of storage, and the power exchange between countries.

Exogenous model inputs include techno-economic parameters such as investment and variable costs, the time series of capacity factors of wind and solar PV, and electricity demand. Electricity demand is assumed to be price-inelastic. To ensure the relevance of our results, we impose certain bounds on the investments of some generation technologies. In particular, we consider the installed storage energy and power capacities of different types of hydropower plants (run-of-river, reservoir, pumped-hydro) and the installed generation capacity of bioenergy to be exogenous, without any possibility of additional investments. Accordingly, there is no need to additionally cap the yearly electricity generation of bioenergy.  Only a subset of countries can install offshore wind power. In Section~\ref{sec: appendix assumptions and data}, we provide more details on assumptions and the input data.

Model results can be interpreted as the outcomes of an idealized, frictionless central European electricity market in which all generators maximize their profits. Real-world market outcomes may differ from this benchmark because of various frictions, i.e., limited information of market actors or barriers to market entry. Note that single countries do not possess individual objective functions, but costs are minimized for the overall interconnected power sector.

For robustness, we do not perform our analysis only for a single weather year, only but for ten different ones covering nearly three decades, i.e.,~1989, 1992, 1995, 1998, 2001, 2004, 2007, 2010, 2013, and 2016. Between these weather years, the time series of renewables, load, and hydro inflow time series differ.

\newpage

% ------------------------------------------
\section{Acknowledgments}
% ------------------------------------------

We thank two anonymous reviewers, the members of the research group \textit{Transformation of the Energy Economy} at DIW Berlin and participants of the \textit{PhD Strommarkttreffen 2021}, the \textit{Enerday 2021}, \textit{IAEE Online Conference 2021}, the \textit{DIW GC Workshop 2021}, the \textit{FSR Summer School 2021}, and the \textit{YEEES 2022 Copenhagen} for very helpful comments and feedback. We gratefully acknowledge financial support from the German Federal Ministry of Economic Affairs and Climate Action (BMWK) via the project MODEZEEN (FKZ 03EI1019D).

% ------------------------------------------
\subsection{Author contributions}
% ------------------------------------------

\textbf{Alexander~Roth:} Conceptualization, Methodology, Software, Formal analysis, Investigation, Data Curation, Writing - Original Draft, Visualization. \textbf{Wolf-Peter~Schill:} Conceptualization, Methodology, Investigation, Writing - Original Draft, Funding acquisition.

% ------------------------------------------
\subsection{Declaration of interests}
% ------------------------------------------

The authors declare no competing interests.

% ------------------------------------------
\subsection{Inclusion and diversity}
% ------------------------------------------

We support inclusive, diverse, and equitable conduct of research. 

% ------------------------------------------
% BIBLIOGRAPHY
% ------------------------------------------

\newpage

\printbibliography

%\putbib

% ------------------------------------------

%\end{bibunit}
\end{refsection}

% ------------------------------------------

% ------------------------------------------
% SUPPLEMENTAL INFORMATION
% ------------------------------------------

\begin{refsection}

\newpage

%\begin{bibunit}
   
% ------------------------------------------

\appendix

\renewcommand{\thesection}{SI}

\pagenumbering{Roman}

\renewcommand{\thetable}{SI.\arabic{table}}
\renewcommand{\thefigure}{SI.\arabic{figure}}

\setcounter{table}{0}
\setcounter{figure}{0}
\setcounter{page}{1}

% ------------------------------------------
\section{Supplemental Information} \label{sec: si}
% ------------------------------------------

\subsection{Assumptions and data} \label{sec: appendix assumptions and data}

\subsubsection{Time series}

All time series concerning generation (capacity factors for solar PV, wind on- and offshore, inflow series for hydropower plants) are taken from ENTSO-E's ``Pan-European Climate Database (PECD)'' \supercite{defelice.2020}. The load data is taken from ENTSO-E's ``Mid-term Adequacy Forecast (MAF) 2020''\supercite{entso-e.2020}.

\subsubsection{Techno-economic parameters for technologies with endogenous capacities}

\begin{table}[H]
\resizebox{\textwidth}{!}{%
\begin{tabular}{l|cccc}
\textbf{Technology} & \textbf{Thermal efficiency {[}\%{]}} & \textbf{Overnight investment costs {[}EUR/kW{]}} & \textbf{Technical Lifetime {[}years{]}} \\ \hline
Bioenergy     & 0.487  & 1951    & 30    \\
Run-of-river  & 0.9    & 600     & 25    \\
PV            & 1      & 3000    & 50    \\
Wind offshore & 1      & 2,506   & 25    \\
Wind onshore  & 1      & 1,182   & 25                                        
\end{tabular}%
}
\caption{Technical and costs assumptions of installable generation technologies (related to STAR Methods)}
\label{tab: assumption generation technologies}
\end{table}

\begin{table}[H]
\resizebox{\textwidth}{!}{%
\begin{tabular}{l|cccccccccc}
\textbf{Technology} & \textbf{\begin{tabular}[c]{@{}c@{}}Marginal costs \\ of storing in \\ {[}EUR/MW{]}\end{tabular}} & \textbf{\begin{tabular}[c]{@{}c@{}}Marginal costs \\ of storing out \\ {[}EUR/MW{]}\end{tabular}} & \textbf{\begin{tabular}[c]{@{}c@{}}Efficiency\\ storing in \\ {[}\%{]}\end{tabular}} & \textbf{\begin{tabular}[c]{@{}c@{}}Efficiency \\ storing out \\ {[}\%{]}\end{tabular}} & \textbf{\begin{tabular}[c]{@{}c@{}}Efficiency \\ self-discharge \\ {[}\%{]}\end{tabular}} & \textbf{\begin{tabular}[c]{@{}c@{}}Overnight \\ investment costs \\ in energy \\ {[}EUR/kWh{]}\end{tabular}} & \textbf{\begin{tabular}[c]{@{}c@{}}Overnight \\ investment costs \\ in capacity charge \\ {[}EUR/kW{]}\end{tabular}} & \textbf{\begin{tabular}[c]{@{}c@{}}Overnight \\ investment costs \\ in capacity discharge \\ {[}EUR/kW{]}\end{tabular}} & \textbf{\begin{tabular}[c]{@{}c@{}}Technical \\ lifetime\\ {[}years{]}\end{tabular}}  \\ \hline
Lithium-Ion & 0.5 & 0.5 & 92 & 92 & 100 & 200 & 150 & 150 & 13 \\
Power-to-gas-to-power & 0.5 & 0.5 & 50 & 50 & 100 & 1 & 3000 & 3000 & 20 \\
Pumped-hydro & 0.5 & 0.5 & 80 & 80 & 100 & 80 & 1100 & 1100 & 60 \\
Reservoir & - & 0.1 & - & 95 & 100 & 10 & - & 200 & 50
\end{tabular}%
}
\caption{Technical and cost assumptions of installable storage technologies (related to STAR Methods)}
\label{tab: assumption storage technologies}
\end{table}

For the principal technical and cost parameters, we rely on previous research\supercite{gaete-morales.2021}, and these are shown in Tables \ref{tab: assumption generation technologies} and \ref{tab: assumption storage technologies}. For all technologies (generation and storage), we assume an interest rate for calculating investment annuities of 4\%. The assumed power of installed bioenergy capacities is provided by ENTSO-E\supercite{entso-e.2018}.

\subsubsection{Exogenous generation and storage capacities}

\begin{table}[H]
\resizebox{\textwidth}{!}{%
\begin{tabular}{ll|cccccccccccc}
\textbf{Technology} & \textbf{Variable} & \textbf{AT} & \textbf{BE} & \textbf{CH} & \textbf{CZ} & \textbf{DE} & \textbf{DK} & \textbf{ES} & \textbf{FR} & \textbf{IT} & \textbf{NL} & \textbf{PL} & \textbf{PT} \\ \hline
Bioenergy & Power {[}GW{]} & 0.50 & 0.62 & 0 & 0.40 & 7.75 & 1.72 & 0.51 & 1.93 & 1.54 & 0.46 & 0.85 & 0.61 \\ \cline{1-2}
Run-of-River & Power {[}GW{]} & 5.56 & 0.17 & 0.64 & 0.33 & 3.99 & 0.01 & 1.16 & 10.96 & 10.65 & 0.04 & 0.44 & 2.86 \\ \cline{1-2}
\multirow{3}{*}{Pumped-hydro (closed)} & Discharging power {[}GW{]} & 0 & 1.31 & 3.99 & 0.69 & 6.06 & 0 & 3.33 & 1.96 & 4.01 & 0 & 1.32 & 0 \\
 & Charging power {[}GW{]} & 0 & 1.15 & 3.94 & 0.65 & 6.07 & 0 & 3.14 & 1.95 & 4.07 & 0 & 1.49 & 0 \\
 & Energy {[}GWh{]} & 0 & 5.30 & 670 & 3.70 & 355 & 0 & 95.40 & 10 & 22.40 & 0 & 6.34 & 0 \\ \cline{1-2}
\multirow{3}{*}{Pumped-hydro (open)} & Discharging power {[}GW{]} & 3.46 & 0 & 0 & 0.47 & 1.64 & 0 & 2.68 & 1.85 & 3.57 & 0 & 0.18 & 2.95 \\
 & Charging power {[}GW{]} & 2.56 & 0 & 0 & 0.44 & 1.36 & 0 & 2.42 & 1.85 & 2.34 & 0 & 0.17 & 2.70 \\
 & Energy {[}GWh{]} & 1722 & 0 & 0 & 2 & 417 & 0 & 6185 & 90 & 382 & 0 & 2 & 1966 \\ \cline{1-2}
\multirow{2}{*}{Reservoir} & Discharging power {[}GW{]} & 2.43 & 0 & 8.15 & 0.70 & 1.30 & 0 & 10.97 & 8.48 & 9.96 & 0 & 0.18 & 3.49 \\
 & Energy {[}GWh{]} & 762 & 0 & 8155 & 3 & 258 & 0 & 11840 & 10000 & 5649 & 0 & 1 & 1187
\end{tabular}%
}
\caption{Assumptions on exogenous generation and storage capacities (related to STAR Methods)}
\label{tab: assumption installed capacities}
\end{table}

\subsubsection{Interconnection capacities}

\begin{table}[H]
\centering
\begin{tabular}{c|c}
\textbf{link} & \multicolumn{1}{c}{\textbf{\begin{tabular}[c]{@{}c@{}}Installed capacity \\ {[}MW{]}\end{tabular}}} \\ \hline
AT\_CH & 1700 \\
AT\_CZ & 1100 \\
AT\_DE & 7500 \\
AT\_IT & 1470 \\
BE\_DE & 1000 \\
BE\_FR & 5050 \\
BE\_NL & 4900 \\
CH\_DE & 5300 \\
CH\_FR & 4000 \\
CH\_IT & 4850 \\
CZ\_DE & 2300 \\
CZ\_PL & 700 \\
DE\_DK & 4000 \\
DE\_FR & 4800 \\
DE\_NL & 5000 \\
DE\_PL & 3750 \\
DK\_PL & 500 \\
ES\_FR & 9000 \\
ES\_PT & 4350 \\
FR\_IT & 3255
\end{tabular}
\caption{Installed Net Transfer Capacities (NTC) in model runs with interconnection (related to STAR Methods)}
\label{tab: assumption installed ntc capacities}
\end{table}

The assumed Net Transfer Capacities (NTC) provided in Table~\ref{tab: assumption installed ntc capacities} are taken from from the TYNDP 2018 (Appendix IV - Cross-border capacities, NTC ST 2040)\supercite{entso-e.2018}.

\begin{figure}[H]
    \centering
    \frame{\includegraphics[width=.65\textwidth]{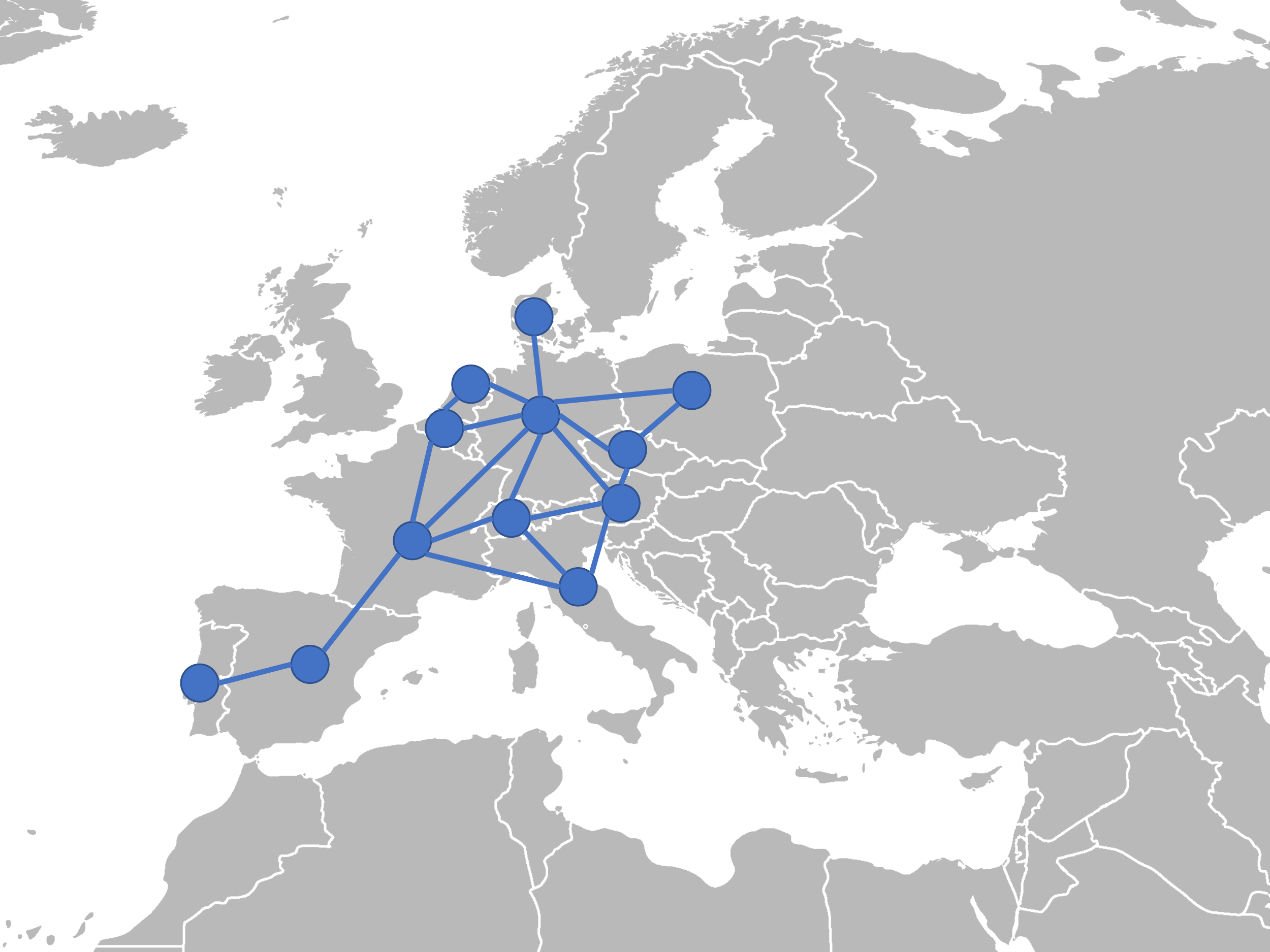}}
    \caption{Geographic scope of the model and existing interconnections (related to STAR Methods)}
    \label{fig: countries}
\end{figure}

Figure \ref{fig: countries} depicts the countries that are part of the model and the respective interconnections between them.

\newpage

\subsection{Model}

Our analysis is model-based, using the open-source capacity expansion model DIETER. A short introduction is provided in section \ref{sec: model}, more details are provided in previous publications\supercite{zerrahn.2017,gaete-morales.2021}.

For illustrative reasons, we provide below the formulation of two key equations of the model: the objective function and the energy balance. Before, we provide a non-exhaustive nomenclature of sets, variables, and parameters used in these equations. Variables are defined with uppercase letters and parameters with lowercase letters.

\paragraph{Sets} $n$ is the set of countries, $h$ the set hours, $dis$ the set of dispatchable generators, $nd$ the set of non-dispatchable generators, and $sto$ the set of storage technologies.

\paragraph{Electricity generation and flows {[}MWh{]}} $G_{n,dis,h}$ is the generation of the dispatchable generation technology $dis$ in country $n$ in hour $h$. $STO^{out}$ is the electricity generation (by discharge) of storage technologies, $STO^{in}$ is the charging, and $RSV^{out}$ is the electricity generation (by outflows) of reservoirs. $F_{l,h}$ is the electric energy sent over line $l$ in hour $h$.

\paragraph{Installed generation capacities {[}MW{]}} $N$ is the installed capacity of a generation technology. $N^{p-out}$ is the installed discharging capacity of storage technologies, $N^{p-in}$ is the installed charging capacity of storage technologies.

\paragraph{Energy installation variables {[}MWh{]}} $N^e$ is the installed energy capacity of storage technologies.

\paragraph{Costs {[}Euro/MW(h){]}} $c^m$ are marginal costs of generation, $c^i$ annualized investment costs of installation power and energy capacities (generation and storage), $c^{fix}$ are the respective annual fixed costs.

\paragraph{Objective function}

DIETER minimizes the total cost~$Z$, consisting of variable generation costs (first term), investment costs of dispatchable and non-dispatchable generators (second term), as well as fixed and variable costs of storage (third term). The objective function of the model is given as:

\begin{align}
Z &= \sum_{n}\Bigg[\sum_{h}\left[\sum_{dis} c^m_{n,dis}G_{n,dis,h} + \sum_{sto}c^{m}_{n,sto} \left(STO^{out}_{n,sto,h} + STO^{in}_{n,sto,h} \right)\right. \nonumber \\ 
& + \left. c^m_{n,rsv} RSV^{out}_{n,rsv,h} \right] \nonumber \\
& + \sum_{dis}\left[\left(c^i_{n,dis} + c^{fix}_{n,dis}\right)N_{n,dis}\right] + \sum_{nd}\left[\left(c^i_{n,nd} + c^{fix}_{n,nd}\right)N_{n,nd}\right] \nonumber \\
& + \sum_{sto}\left[\left(c^{i,p-out}_{n,sto} + c^{fix,p-out}_{n,sto}\right)N^{p-out}_{n,sto} + \left(c^{i,p-in}_{n,sto} + c^{fix,p-in}_{n,sto}\right)N^{p-in}_{n,sto} \right. \nonumber \\ 
& + \left. \left(c^{i,e}_{n,sto} + c^{fix,e}_{n,sto}\right)N^{e}_{n,sto}\right] \Bigg] \label{eq:objective}
\end{align}

Those fixed variables (NTC capacities, installed capacities of hydro and bioenergy), and some nomenclature details, are omitted in the objective function for the reader's convenience. The full objective function is provided in the model code.

\paragraph{Energy balance}

The wholesale energy balance reads as follows:

\begin{align}
& d_{n,h} + \sum_{sto} STO^{in}_{n,sto,h} \nonumber \\
& = \nonumber \\
& \quad \sum_{dis} G_{n,dis,h} + \sum_{nd}G_{n,nd,h} + \sum_{sto}STO^{out}_{n,sto,h} + \sum_{rsv}RSV^{out}_{n,rsv,h} \nonumber \\ 
& + \sum_{l}i_{l,n}F_{l,h} \qquad \forall n,h 
\label{eq:energy_balance}
\end{align}

The left-hand side is total electricity demand in hour~$h$ at node~$n$ plus charging of storage technologies; the right-hand side is the total generation, including storage discharging, plus net imports:~$F_{l,h}$ represents the directed flow on line~$l$. If~$F_{l,n}>0$, electricity flows from the source to the sink of the line and reversed for~$F_{l,n} < 0$. With the incidence parameter~$i_{l,n}\in\lbrace-1,0,1\rbrace$, source, and sink are exogenously defined. 

\subsection{Background on factorization} \label{sec: appendix factorization}

To identify the importance of different factors that reduce optimal storage need through interconnection, we (1) define several counterfactual scenarios and (2) then attribute the overall change to different factors using a ``factorization'' method\supercite{stein.1993,lunt.2021}.

To explain the principles of factorization, we borrow an example used in another paper\supercite{lunt.2021}. Using a case study from the field of climate science, we aim to explain why oceans around 3 million years ago were  warmer than today. Assuming that two important factors are atmospheric CO$_2$ concentration and the extent and volume of large ice sheets, we apply a climate model and run several counterfactual scenarios. Both factors can have two kinds of states: CO$_2$ concentration can be low or high, and ice sheets can be small or large. Comparing different model outcomes, we can identify a ``sole'' CO$_2$ and ice sheet effect, but also an interaction effect between CO$_2$ concentration and ice sheet extension on ocean temperature.

Following the notation introduced in previous research\supercite{schar.2017}, we describe the different scenarios in the following way: in $f_0$, ice sheets are small, and CO$_2$ is low. In the scenario $f_1$, the ice sheets are large, but CO$_2$ concentration is low. In scenario $f_2$, ice sheets are small, but CO$_2$ concentration is high. Finally, in scenario $f_{12}$, ice sheets are large, and CO$_2$ concentration is high. 

The factorization method on which we rely on\supercite{stein.1993} defines the impact of the different factors in the following way:
\begin{align}
\hat{f}_1    & = f_{1} - f_{0}, \label{eq1_1} \\
\hat{f}_2    & = f_{2} - f_{0}. \label{eq1_2}
\end{align}
$\hat{f}_1$ is the sole contribution of ice sheets, $\hat{f}_2$ of CO$_2$ concentration to the change in ocean temperature. However, with this factorization approach, the sum of the individual effects does not (in general) add up to the overall effect:
\begin{equation}
    \hat{f}_1 + \hat{f}_2  \neq f_{12} - f_{0}
\end{equation}
Thus, an ``interaction effect'' $\hat{f}_{12}$ is introduced, which captures the joint effect of ice sheets size and CO$_2$ concentration on ocean temperature\supercite{stein.1993}, such that $\hat{f}_1$, $\hat{f}_2$, and $\hat{f}_{12}$ add up to total the total effect $f_{12}-f_0$:
\begin{align}
    f_{12} - f_0 = & \hat{f}_1 + \hat{f}_2 + \hat{f}_{12}  \nonumber \\
    \Leftrightarrow  \qquad \qquad \hat{f}_{12} = & f_{12} - f_0 - \hat{f}_1 - \hat{f}_2  \nonumber \\
    \Leftrightarrow  \qquad \qquad \hat{f}_{12} = & f_{12} - f_0 - (f_1 -f_0) - (f_2 -f_0)  \nonumber \\
    \Leftrightarrow  \qquad \qquad \hat{f}_{12} = & f_{12} - f_{1} - f_{2} + f_0
\end{align}
If interested in the overall effect of CO$_2$ concentration and ice sheets on ocean temperatures and not in the interaction term, $\hat{f}_{12}$ has to be ``distributed'' to the other factors $\hat{f}_{1}$ and $\hat{f}_{2}$. This distribution can be done in different ways. One possibility is to share that interaction term equally between the two factors that are involved in that interaction term. Following that logic, the total effect of the two factors becomes:
\begin{align}
\hat{f}_1^{total} & = f_{1} - f_{0} + \frac{1}{2}\hat{f}_{12} = \frac{1}{2}((f_1 - f_0)+(f_{12}-f{2})) \\
\hat{f}_2^{total} & = f_{2} - f_{0} + \frac{1}{2}\hat{f}_{12} = \frac{1}{2}((f_2 - f_0)+(f_{12}-f{1}))
\end{align}
and capture the overall effect of ice sheets ($\hat{f}_1$) and CO$_2$ concentration ($\hat{f}_2$) on ocean temperatures. For a complete decomposition of factors, $2^n$ runs have to be conducted where $n$ is the number of factors.

\subsection{Overview of scenario runs}

\begin{table}[H]
\resizebox{\textwidth}{!}{%
\begin{tabular}{l|l|l|l|l|l|l|l}

\textbf{Run} & \textbf{Identifier} & \textbf{(1) Interconnection} & \textbf{(2) Wind} & \textbf{(3) PV} & \textbf{(4) Load} & \textbf{(5) Hydro} & \textbf{(6) Bio}        \\ \hline
1     & $f_0$        & no      & harmonized     & harmonized     & harmonized     & harmonized     & harmonized     \\ 
2     & $f_1$        & yes     & harmonized     & harmonized     & harmonized     & harmonized     & harmonized     \\ 
3     & $f_2$        & no      & not harmonized & harmonized     & harmonized     & harmonized     & harmonized     \\ 
4     & $f_3$        & no      & harmonized     & not harmonized & harmonized     & harmonized     & harmonized     \\ 
5     & $f_4$        & no      & harmonized     & harmonized     & not harmonized & harmonized     & harmonized     \\ 
6     & $f_5$        & no      & harmonized     & harmonized     & harmonized     & not harmonized & harmonized     \\ 
7     & $f_6$        & no      & harmonized     & harmonized     & harmonized     & harmonized     & not harmonized \\ 
8     & $f_{12}$     & yes     & not harmonized & harmonized     & harmonized     & harmonized     & harmonized     \\ 
9     & $f_{13}$     & yes     & harmonized     & not harmonized & harmonized     & harmonized     & harmonized     \\ 
\dots & \dots        & \dots   & \dots          & \dots          & \dots          & \dots          & \dots          \\ 
63    & $f_{23456}$  & no      & not harmonized & not harmonized & not harmonized & not harmonized & not harmonized \\ 
64    & $f_{123456}$ & yes     & not harmonized & not harmonized & not harmonized & not harmonized & not harmonized \\ 

\end{tabular}%
}
\caption{Overview of scenario runs (related to STAR Methods)}
\label{tab: appendix overview runs}
\end{table}

Table \ref{tab: appendix overview runs} provides an intuition of which scenario runs are performed and how they are defined. For every weather year, 64 runs are needed for a complete factorization.

\subsection{Further results} \label{sec: si further results}

\begin{figure}[H]

\centering
     \begin{subfigure}[c]{0.45\textwidth}
         \centering
         \subcaption{Storage energy, short duration}
         \includegraphics[width=\textwidth]{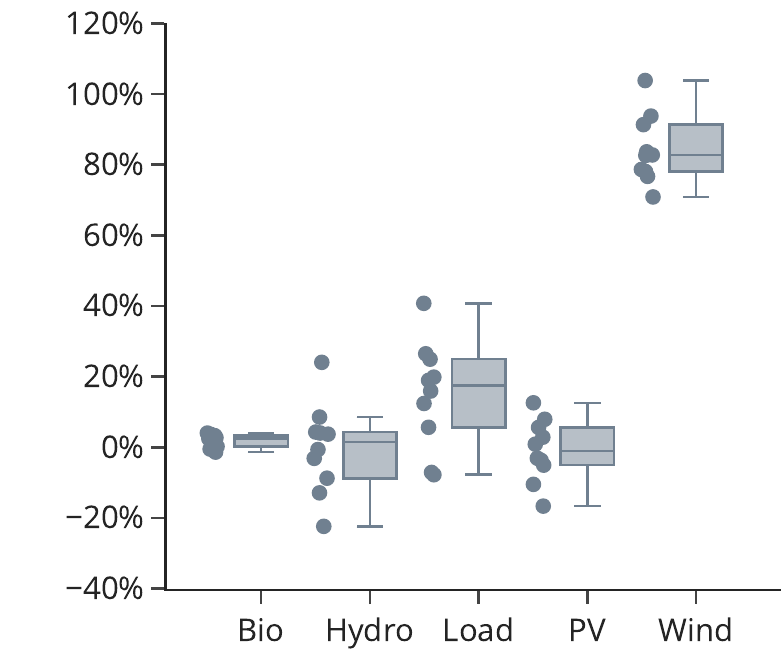}
     \end{subfigure}
     \begin{subfigure}[c]{0.45\textwidth}
         \centering
         \subcaption{Storage energy, long duration}
         \includegraphics[width=\textwidth]{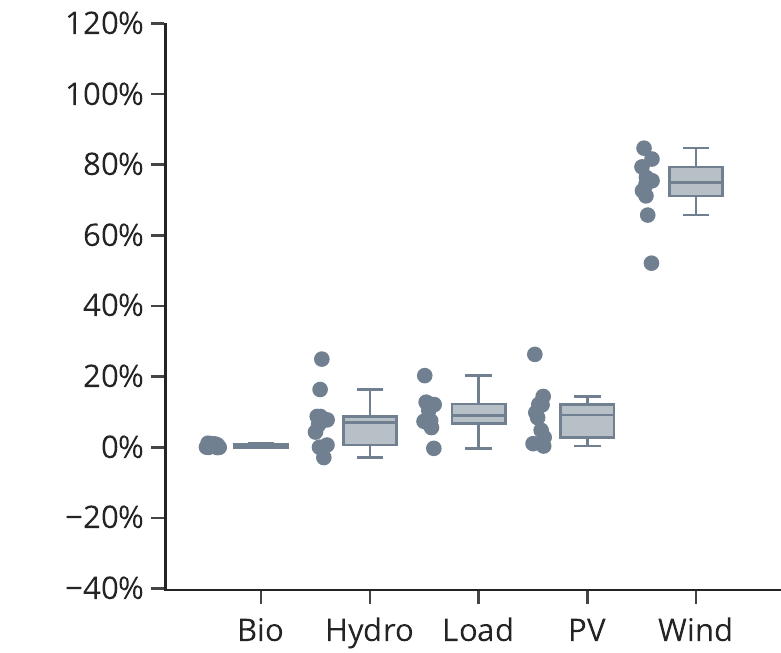}
    \end{subfigure}
    \vspace{2mm}
    
    \begin{subfigure}[c]{0.45\textwidth}
         \centering
         \subcaption{Storage discharging power, short duration}
         \includegraphics[width=\textwidth]{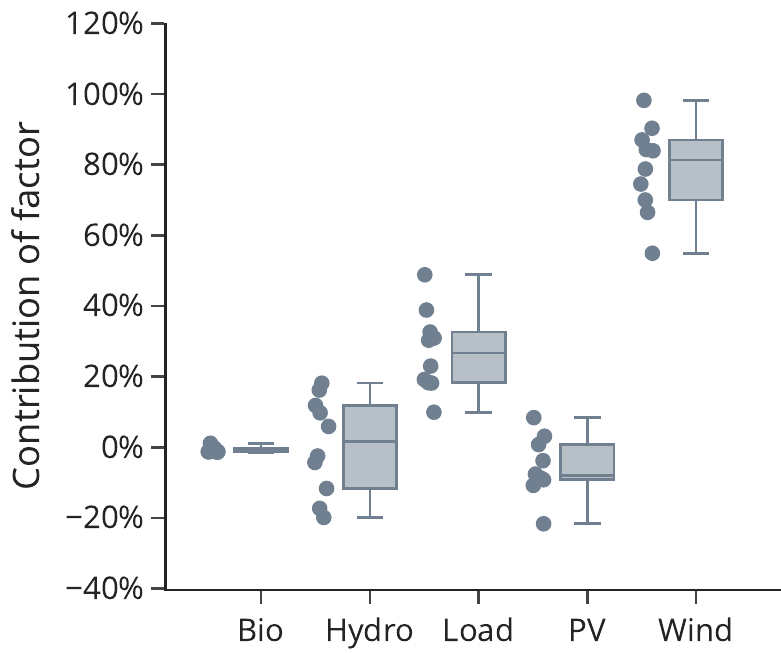}
    \end{subfigure}
    \begin{subfigure}[c]{0.45\textwidth}
        \centering
        \subcaption{Storage discharging power, long duration}
        \includegraphics[width=\textwidth]{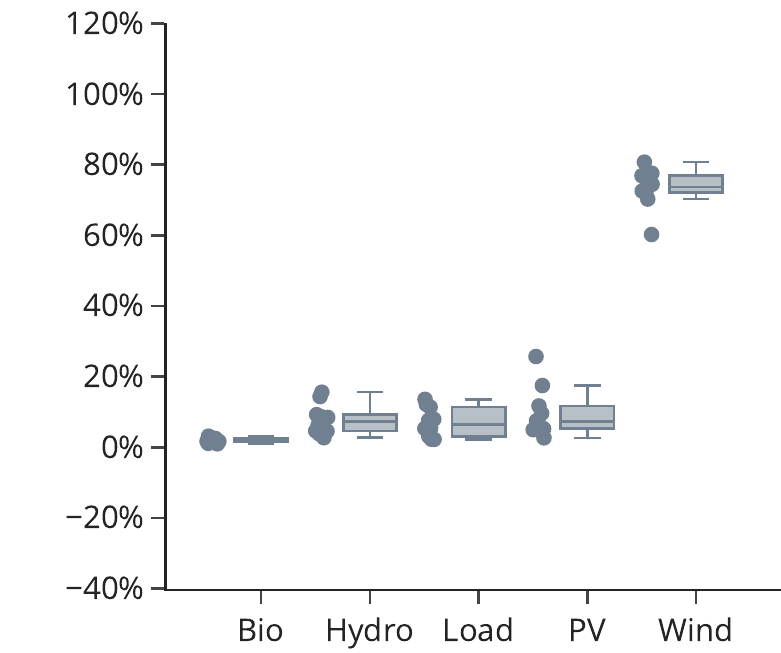}
    \end{subfigure}
    \vspace{2mm}

     \begin{minipage}[c]{0.90\textwidth}
     \medskip\footnotesize
        \emph{Notes:} Every dot is the scenario result based on one weather year. The middle bar shows the median value. The box shows the interquartile range (IQR), which are all values between the \nth{1} and \nth{3} quartile. The whiskers show the range of values beyond the IQR, with a maximum of 1,5 x IQR below the \nth{1} quartile and above the \nth{3} quartile.
     \end{minipage}
     
     \caption{Relative contribution of different factors to the change in storage energy and discharging power capacity related to interconnection (related to Figure \ref{fig: waterfall storage})}
     \label{fig: percentage change storage}
     
\end{figure}

Heterogeneity in wind power explains between 55\% and 104\% of short-duration storage energy and discharging power capacity reduction and 52\% to 85\% of long-duration storage capacity reductions, respectively. At the other end of the spectrum, country-specific differences in installed bioenergy hardly have an effect. Differences in hydropower, load time series, and PV profiles have varying contributions, especially for short-duration storage. The effect of hydropower ranges between -22\% and +24\% for storage energy and -20\% and +18\% for storage discharging power (Figure \ref{fig: percentage change storage}).

We find similar outcomes for PV. The effect of different PV capacity factors through interconnection on aggregate optimal short-duration storage energy or discharging capacity varies between -17\% and +13\%, or -22\% and 8\%, respectively. This contrasts with the results for wind power, which always decreases storage needs. 

Negative percentage values indicate that the current heterogeneous mix of hydro capacities (run-of-river, reservoirs, and pumped hydro) may even increase optimal storage needs compared to a setting with equal relative shares, thus harmonized installations. Exploring this combined technology effects in detail merits further investigation.

Overall, the influence of different weather years on the composition of the factors is more pronounced for short-duration than for long-duration storage.

\begin{figure}[H]
    \centering
        \includegraphics[width=.75\textwidth]{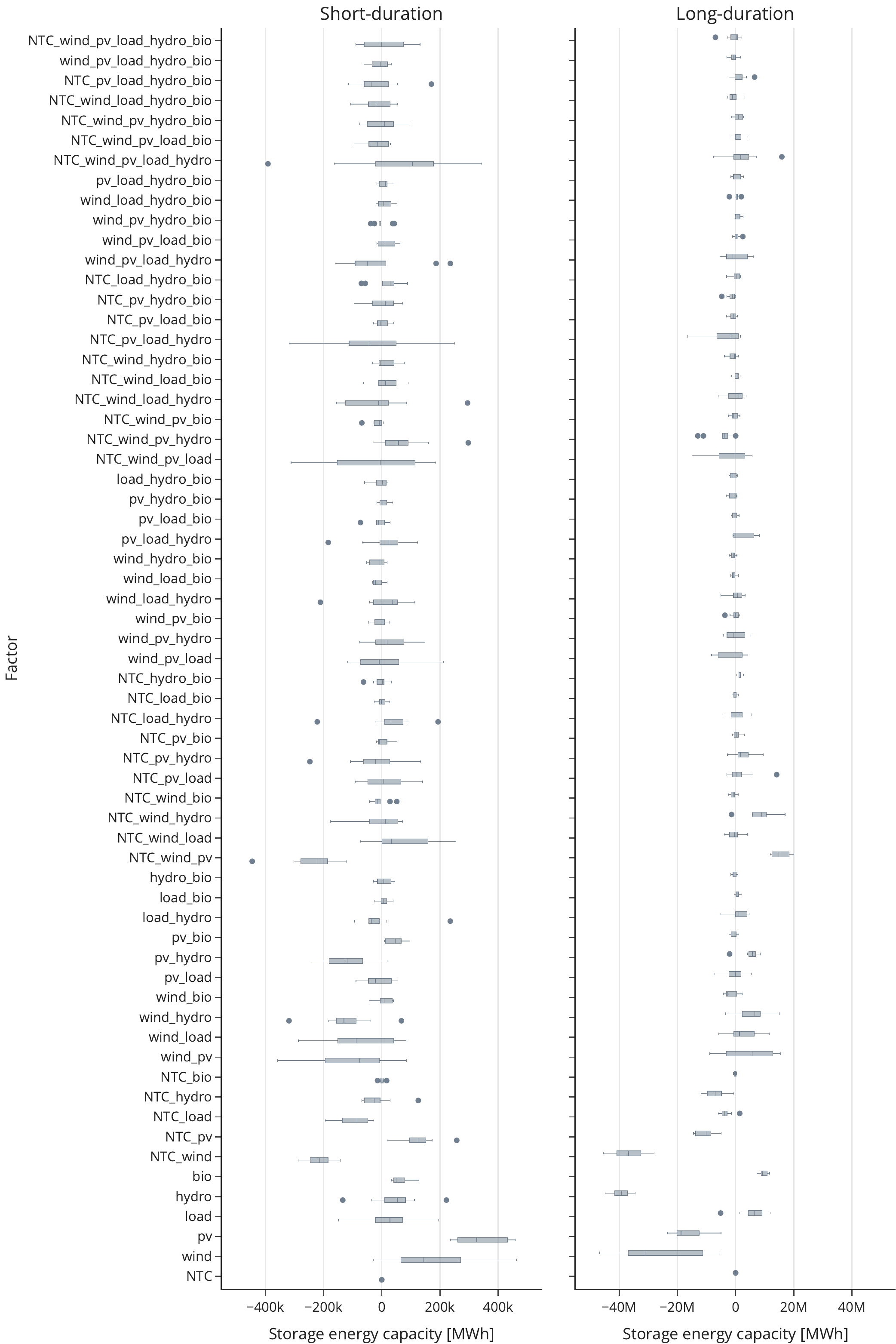}
    \caption{Impact of all factors on storage energy capacity in all years (related to Figure \ref{fig: waterfall storage})}
    \label{fig: all factors energy all years}
    
    \vspace{2mm}
     \begin{minipage}[c]{0.85\textwidth}
     \medskip\footnotesize
        \emph{Notes:} Differentiated by short- and long-duration, the strength of each individual factor is depicted, covering all 10 weather years. If below zero, a factor negatively impacts aggregate optimal storage energy capacity. If above zero, a factor increases aggregate optimal storage energy capacity.
     \end{minipage}
\end{figure}

\begin{figure}[H]
    \centering
    \includegraphics[width=.75\textwidth]{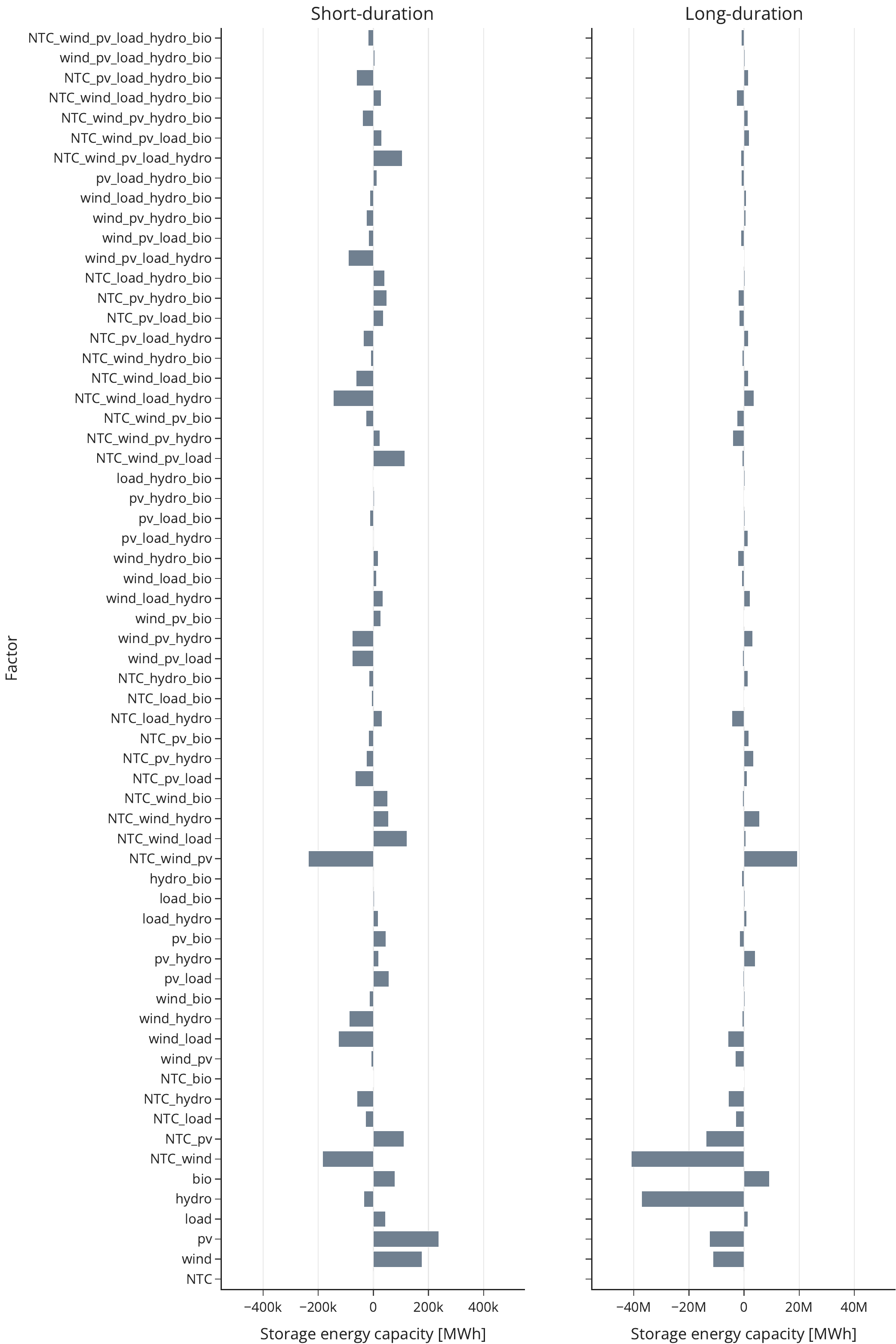}
    \caption{Impact of all factors on storage energy capacity in 2016 (related to Figure \ref{fig: waterfall storage})}
    \label{fig: all factors energy 2016}
    \vspace{2mm}
     \begin{minipage}[c]{0.85\textwidth}
     \medskip\footnotesize
        \emph{Notes:} Differentiated by short- and long-duration, the strength of each individual factor is depicted for the weather year 2016. If below zero, a factor negatively impacts aggregate optimal storage energy capacity. If above zero, a factor increases aggregate optimal storage energy capacity.
     \end{minipage}
\end{figure}

\begin{figure}[H]

\centering
     \begin{subfigure}[c]{0.66\textwidth}
         \centering
         \subcaption{Relative, per country}
         \includegraphics[width=\textwidth]{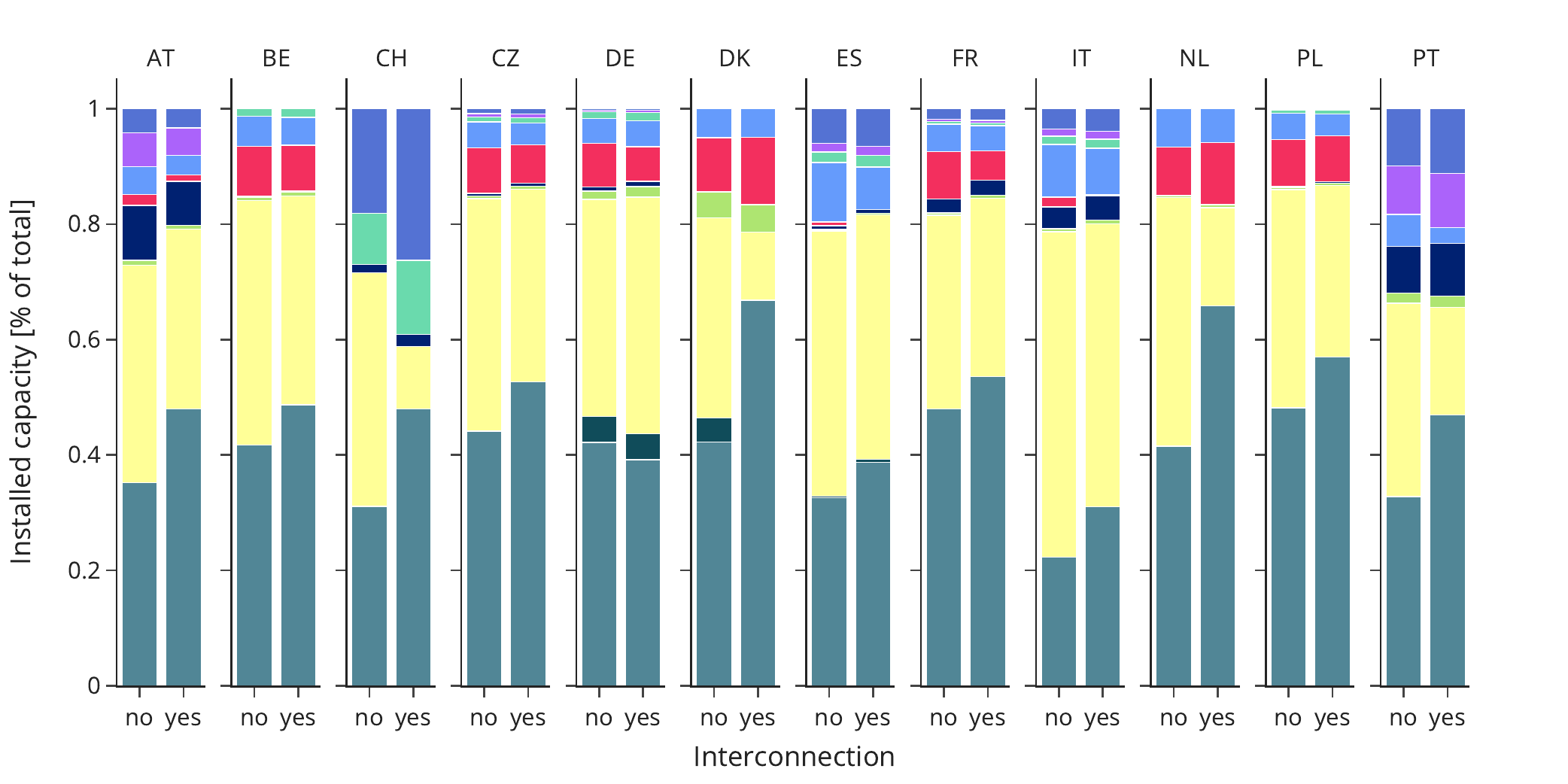}
     \end{subfigure}
     \begin{subfigure}[c]{0.33\textwidth}
         \centering
         \subcaption{Absolute, overall}
         \includegraphics[width=\textwidth]{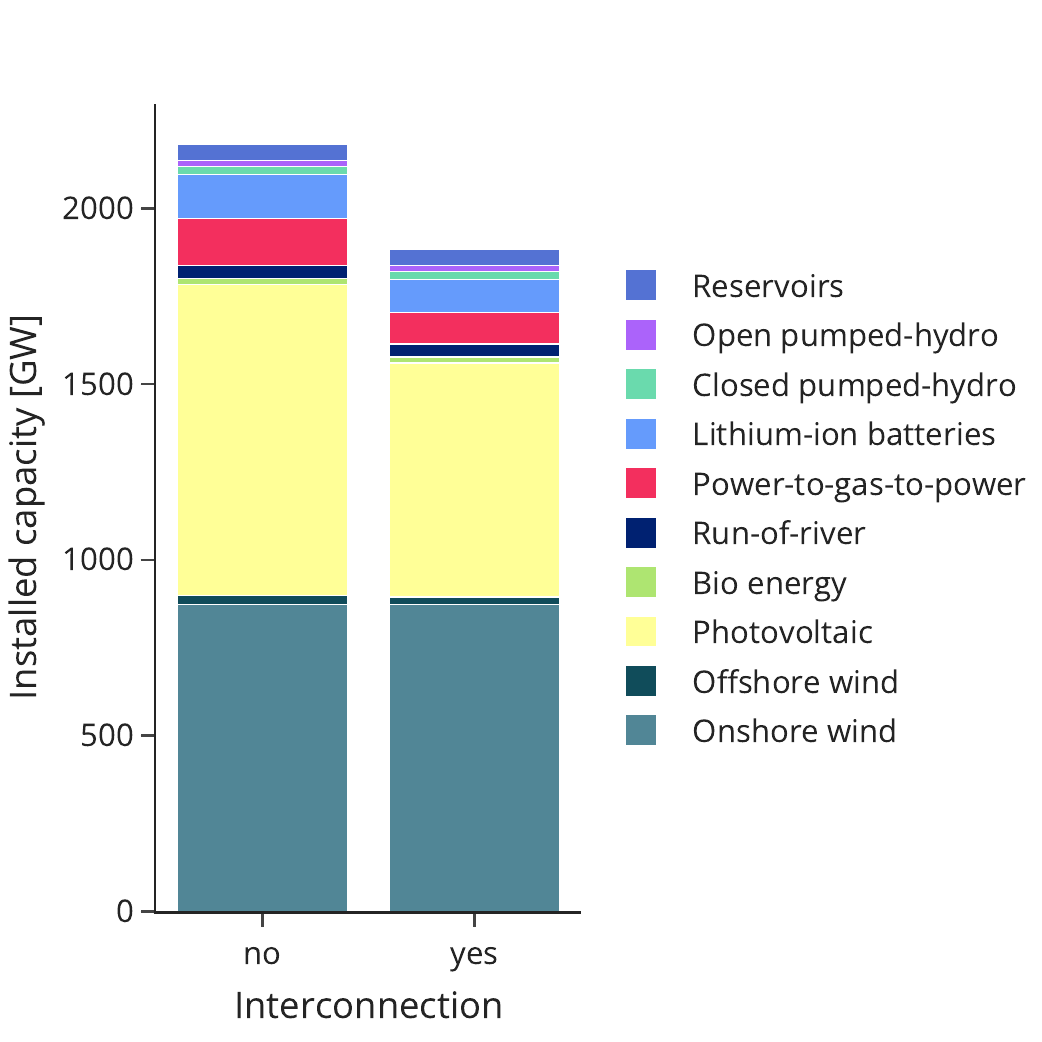}
    \end{subfigure}
    
%    \vspace{2mm}
%
%     \begin{minipage}[c]{0.90\textwidth}
%     \medskip\footnotesize
%        \emph{Notes:} 
%     \end{minipage}
     
     \caption{Installed power plant and storage discharging capacities for scenarios with or without interconnection (related to Figure \ref{fig: waterfall storage})}
     \label{fig: power plant split}
     
\end{figure}

Figure~\ref{fig: power plant split} shows optimal generation and storage capacities for scenarios with and without interconnection. While solar PV and onshore wind power dominate the capacity mix in all countries, the share of onshore wind power increases in scenarios with interconnection compared to the setting with isolated power systems in all countries (left panel). This further corroborates our conclusion that geographical balancing particularly helps to smooth wind power variability across countries. The Figure also shows that the overall generation capacity decreases in a setting with interconnection (right panel). This is largely driven by a lower need for solar PV generation capacity, enabled by lower curtailment and better (cross-border) use of installed wind power capacities.

\begin{figure}[H]
\centering
\includegraphics[width=\textwidth]{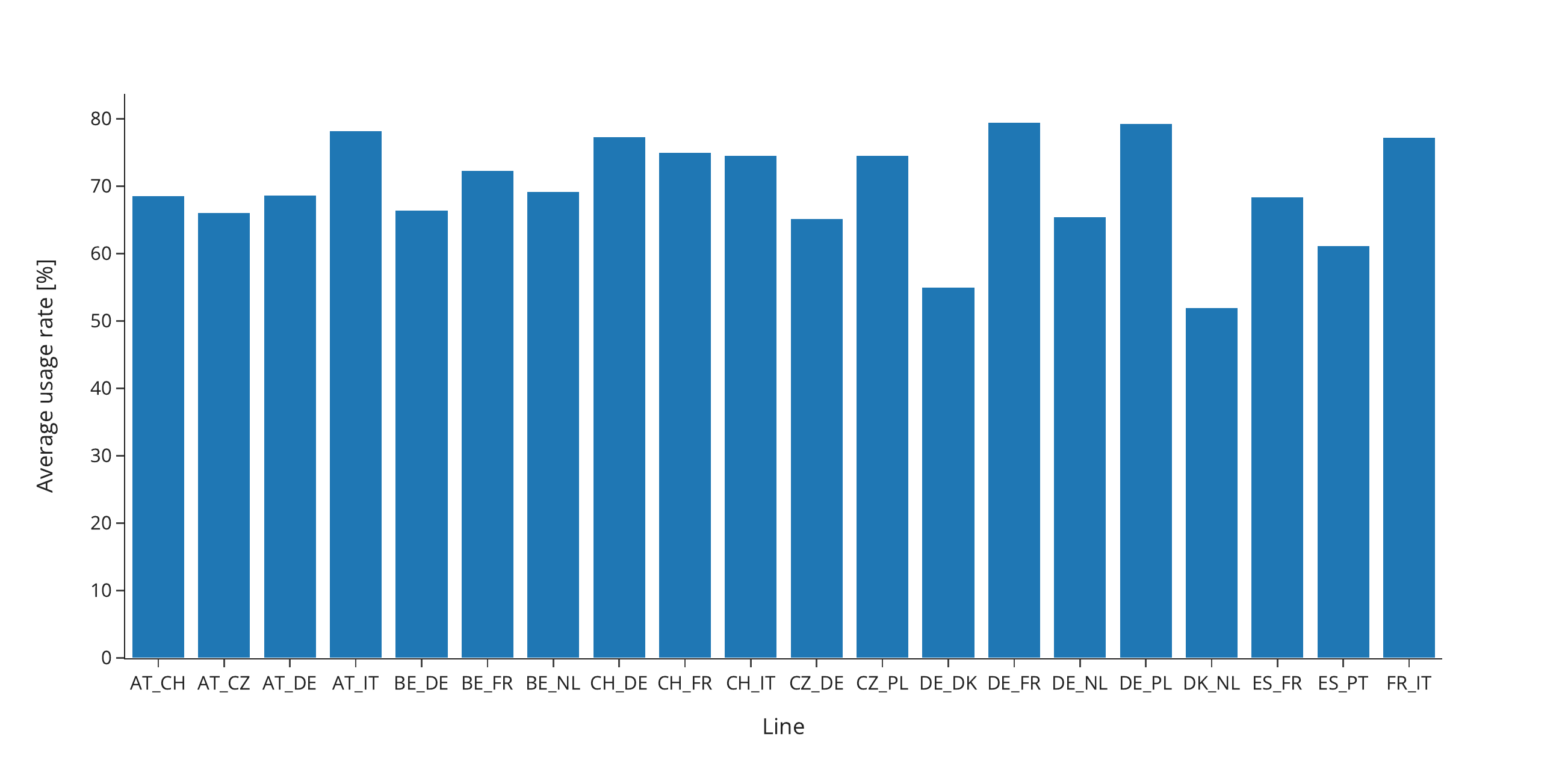}
\vspace{1mm}
\begin{minipage}[c]{0.90\textwidth}
\medskip\footnotesize
    \emph{Notes:} Data of the weather year 2016 shown.
\end{minipage}
\caption{Average hourly usage rates of interconnections (related to STAR Methods)}
\label{fig: interconnection utilization}
\end{figure}

Average utilization rates of the modeled interconnections are both relatively high and homogeneous, with values between around 50\% and 80\% (Figure~\ref{fig: interconnection utilization}). Such high usage rates imply that the NTC expansion assumed by ENTSO-E\supercite{entso-e.2018} for 2040 may not be sufficient for the fully renewable central European power sector modeled here. The connections between Germany and its neighbors France, Poland, and Switzerland, as well as the lines between Austria and Italy and between France and Italy, are most heavily used. This indicates that further extensions of these connections would be particularly desirable. In contrast, interconnections between Denmark and the Netherlands as well as Denmark and Germany are relatively under-utilized. 

% ------------

\newpage

\printbibliography
%\putbib

% ------------

%\end{bibunit}
\end{refsection}

% ------------

\end{document}

% ------------